\documentclass[twocolumn]{aastex701}
\usepackage{float}
\usepackage{caption} 
\usepackage{graphicx}
\usepackage{amsmath}
\newcommand{\msun}{{\rm M_\odot}}

\usepackage{rotating}
\usepackage{xcolor}
\usepackage{soul}

\begin{document}


\title{Connecting the Dots: UV-Bright Companions of Little Red Dots as Lyman-Werner Sources\\Enabling Direct Collapse Black Hole Formation}

\author[orcid=0009-0005-2295-7246,sname='Josephine F.W. Baggen']{Josephine F.W. Baggen}
\affiliation{Department of Astronomy, Yale University, New Haven, CT 06511, USA}
\email[show]{josephine.baggen@yale.edu}
\author[orcid=0000-0002-0748-9115,sname='Matthew T. Scoggins']{Matthew T. Scoggins}
\email{mts2188@columbia.edu}
\affiliation{Department of Astronomy, Columbia University, 550 West 120th Street, New York, NY, 10027, U.S.A}
\author[orcid=0000-0002-8282-9888,sname='Pieter van Dokkum']{Pieter van Dokkum}
\affiliation{Department of Astronomy, Yale University, New Haven, CT 06511, USA}
\email{pieter.vandokkum@yale.edu}
\author[orcid=0000-0003-3633-5403,sname='Zolt\'an Haiman']{Zolt\'an Haiman}
\email{Zoltan.Haiman@ista.ac.at}
\affiliation{Institute of Science and Technology Austria (ISTA), Am Campus 1, 3400 Klosterneuburg, Austria}
\affiliation{Department of Astronomy, Columbia University, 550 West 120th Street, New York, NY, 10027, U.S.A}
\affiliation{Department of Physics, Columbia University, 550 West 120th Street, New York, NY, 10027, U.S.A}
\author[orcid=0000-0001-5586-6950,sname='Alberto Torralba']{Alberto Torralba}
\email{Alberto.Torralba@ista.ac.at}
\affiliation{Institute of Science and Technology Austria (ISTA), Am Campus 1, 3400 Klosterneuburg, Austria}
\author[orcid=0000-0003-2871-127X,sname='Jorryt Matthee']{Jorryt Matthee}
\email{jorryt.matthee@ist.ac.at}
\affiliation{Institute of Science and Technology Austria (ISTA), Am Campus 1, 3400 Klosterneuburg, Austria}





\begin{abstract}
We compile a sample of 83 little red dots (LRDs) with 
JWST imaging and find that a substantial fraction ($\sim$43\%, rising to $\gtrsim$80\% for the most luminous LRDs) host one or more spatially offset, UV-bright companions at projected separations of $0.5$\,kpc\,$\lesssim d\lesssim 5$\,kpc, with median $\langle d \rangle = 1.0$\,kpc.
This fraction is even higher when smaller spatial scales are probed at high signal-to-noise ratio: the two most strongly lensed LRDs, A383-LRD1 and the newly discovered A68-LRD1, both have UV-bright companions at separations of only $d\sim0.3$ kpc, below the resolution limit of most unlensed JWST samples.
We explore whether these ubiquitous red\,/\,blue  configurations may be physically linked to the formation of LRDs, in analogy with the ``synchronized pair" scenario originally proposed for direct-collapse black hole formation. 
In this picture, UV radiation from the companions, with typically modest stellar masses ($M_* \sim 10^{8}-10^9\msun$), 
suppresses molecular hydrogen cooling in nearby gas, allowing nearly isothermal collapse and the formation of extremely compact objects, such as massive black holes, supermassive stars or quasi-stars. Using component-resolved photometry and spectral energy distribution modeling, we infer Lyman-Werner radiation fields of $J_{21,\rm LW}\sim10^{2.5}$-$10^{5}$ at the locations of the red components, comparable to those required in direct-collapse models, suggesting that the necessary photodissociation conditions are realized in many LRD systems. This framework provides a simple and self-consistent explanation for the extreme compactness and distinctive spectral properties of LRDs and links long-standing theoretical models for early compact object formation directly to a population now observed with JWST in the early Universe.
\end{abstract}




\section{Introduction} 
\label{sec:introduction}
The James Webb Space Telescope (JWST) has revealed a population of extremely compact red sources, referred to as little red dots \citep[LRDs;][]{Matthee2024}, predominantly found at high redshifts \citep[$z \sim 4$--$8$; e.g.,][]{Kocevski2025ApJ...986..126K, Pacucci2025_lowspin, Billand2026A&A...706A..29B}. Their unique spectral and morphological
properties challenge simple interpretations in which the observed emission arises from normal stellar populations or classical active galactic nuclei \citep[see][for detailed discussions]{InayoshiHo2025, Matthee2025ConPh..66..116M, Volonteri2025}.

While dubbed ``dots'' owing to their compact appearance
at rest-frame optical wavelengths, LRDs are increasingly revealed to be more complex than single, unresolved point sources, particularly in filters tracing the rest-frame ultraviolet (UV) \citep[e.g.,][]{Baggen2023,Baggen2024, Baggen2025,Labbe2024,Tanaka2024arXiv241214246T_dualLRD, Chen2025,Golubchik2025,Merida2025A&A...698A.317M,
Rinaldi2025ApJ...992...71R_notdot, Rinaldi2025arXiv250717738R, Torralba2026, DEugenio2026, Yanagisawa2026}.
One of the clearest examples of such resolved substructure is observed in A383-LRD1, a strongly lensed system behind the galaxy cluster A383 \citep{Baggen2025,Golubchik2025,Knudsen2025}. Strong gravitational lensing enables the separation of two compact components that are $\sim$300-400~pc apart in the source plane, with one component dominating the rest-frame UV with a flat continuum and the other dominating the rest-frame optical, with a very steep ``BH*/Cliff"-like spectral energy distribution (SED) \citep{Graaff2025_Cliff,Naidu2025_BHstar}. In this case, the characteristic V-shaped SED associated with LRDs \citep[e.g.,][]{Hviding2025,Setton2025ApJ...995..118S_vshape} arises from the superposition of these two components. Without the aid of strong gravitational lensing, these components would be blended at typical JWST resolution, making it difficult to disentangle their distinct contributions to the observed SED \citep{Baggen2025}.

Beyond elucidating the origin of the V-shaped SEDs of LRDs, this configuration raises a more fundamental question: why is an extremely compact red source found in such close proximity to a UV-bright companion? Is this a coincidence, or does it reflect a physical connection that is important for the formation of these enigmatic objects?

Interestingly, theoretical models have long predicted that the formation of extremely compact massive objects in the early Universe requires precisely such environments. 
In nearly metal-free gas, molecular hydrogen provides the dominant cooling channel, but far-UV photons in the Lyman-Werner (LW) band can dissociate H$_2$ via the Solomon process \citep{Draine1996}. 
If the local LW radiation field is sufficiently intense, molecular hydrogen is dissociated, and cooling is suppressed, allowing gas clouds to bypass normal star formation and collapse rapidly at the atomic-cooling threshold~\citep{Omukai2001}, avoiding fragmentation~\citep{Oh2002} and ultimately leading to a massive black hole~\citep{Bromm_Loeb2003,VolonteriRees2005, LodatoNatarajan2006MNRAS.371.1813L, LodatoNatarajan2007MNRAS.377L..64L}.
This has been envisioned to occur through the intermediate formation of a supermassive star \citep{Agarwal_2012, Latif2013MNRAS.433.1607L, Ferrara2014MNRAS.443.2410F, Sugimura_2014,Chon_2016,Hosokawa_2016, Hirano_2017, Haemmerle_2018},
a short-lived quasi-star phase \citep[e.g.,][]{Begelman2006, VolonteriBegelman2010MNRAS.409.1022V}, 
or an extremely dense star cluster \citep[e.g.,][]{Omukai2008,DevecchiVolonteri2009ApJ...694..302D, Boekholt_2018, Natarajan2011BASI...39..145N}.
These pathways can all ultimately seed a massive ``heavy seed" black hole \citep[see][for comprehensive reviews]{Volonteri2010_review,Haiman2013_review, Inayoshi2020ARA&A..58...27I, Volonteri2025}, but they rely on the suppression of efficient ${\rm H_2}$ cooling and fragmentation of a near-pristine cloud.

The strength of the LW radiation field is conventionally quantified by the specific intensity $J_{21}$, expressed in units of $10^{-21}\,\mathrm{erg\,s^{-1}\,cm^{-2}\,Hz^{-1}\,sr^{-1}}$. Molecular cooling is suppressed only when the local LW intensity exceeds a critical threshold, $J_{\rm crit}$, whose value depends on the spectral shape of the radiation field and the thermochemical state of the gas. Estimates of $J_{\rm crit}$ have spanned a wide range \citep[e.g.,][]{Haiman_1997,Machacek2001,Omukai2001,Oh2002, Shang2010,Wolcott-Green_Haiman_Bryan_2011MNRAS.418..838W,Latif2014_uv_flux,Agarwal2016MNRAS.459.4209A,Inayoshi2020ARA&A..58...27I}.
However, in gas clouds in protogalactic halos with virial temperatures above $\sim10^4$K, where atomic cooling is activated, allowing the gas to contract to high densities, studies have converged on $J_{\rm crit} \approx 10^{3}\,J_{21}$ for realistic SEDs \citep{Sugimura_2014,Wolcott-Green2017}.
Such intensities far exceed the global UV background, yet they can be achieved in the immediate vicinity of star-forming galaxies, known as the ``close synchronized pair" scenario \citep{Dijkstra2008, Agarwal2014, Visbal2014, Regan2017}. 

Although originally developed to explain the origin of the seeds of supermassive black holes in quasars at $z \sim 6$, the synchronized pair scenario may be relevant for LRDs as well. 
Recent theoretical work has indeed begun to explore LRDs as the outcome of rapid, fragmentation-suppressed collapse, namely 
as direct-collapse black holes (DCBHs) \citep{Scoggins_2024, Cenci2025, Fei2025arXiv250920452F, Jeon2026ApJ...998..148J, Pacucci2026arXiv260114368P}, successfully reproducing key observed properties \citep{Pacucci2026arXiv260114368P}, 
as well as their expected progenitors, such as supermassive stars and quasi-stars \citep{Fei2025arXiv250920452F,Zwick2025,Begelman2026,Cantiello2026ApJ..1000L...4C,Chisholm2026arXiv260215935C,NandalLoeb2026ApJ...998..124N, Santarelli2026ApJ...998..150S,Santarelli2026ApJ...998L...4S_LRD}. The steep red rest-frame optical spectra, with Balmer absorption features and prominent Balmer breaks, observed in LRDs may reflect the dense gas environments expected after such a collapse \citep{Juodzbalis2024_rosetta,
Graaff2025_Cliff,Graaff2025_unified, 
DEugenio2025,
InayoshiMaiolino2025,
Ji2025,
Kido2025MNRAS.544.3407K,
Liu2025ApJ...994..113L,
Naidu2025_BHstar,
MadauMaiolino2026arXiv260222386M,
Matthee2026arXiv260317667M,
Rusakov2026Natur.649..574R,
Sneppen2026arXiv260118864S}.

If LRDs are indeed the products of this direct-collapse process, 
and have not yet evolved far away from their formation conditions, then they should generally appear in close proximity to UV-bright companions capable of generating $J_{21}$ flux levels exceeding the critical threshold. 
In this Letter, we test this hypothesis across a large spectroscopic sample of LRDs, including both unlensed deep fields and strongly lensed cluster fields. 
We search for nearby UV-bright companions around LRDs and quantify whether their UV output is sufficient to generate intense local LW radiation fields comparable to those required by theoretical models.

Throughout this paper, we adopt the flat Planck18 CDM cosmology \citep[with BAO constraints; Table 2 in][]{Planck2020}, and report all magnitudes in the AB system.

\begin{figure*}
    \centering
    \begin{minipage}{\linewidth}
        \centering
        {\large\textbf{Literature Sample}}\par
        \vspace{0.2cm}
        \includegraphics[width=\linewidth]{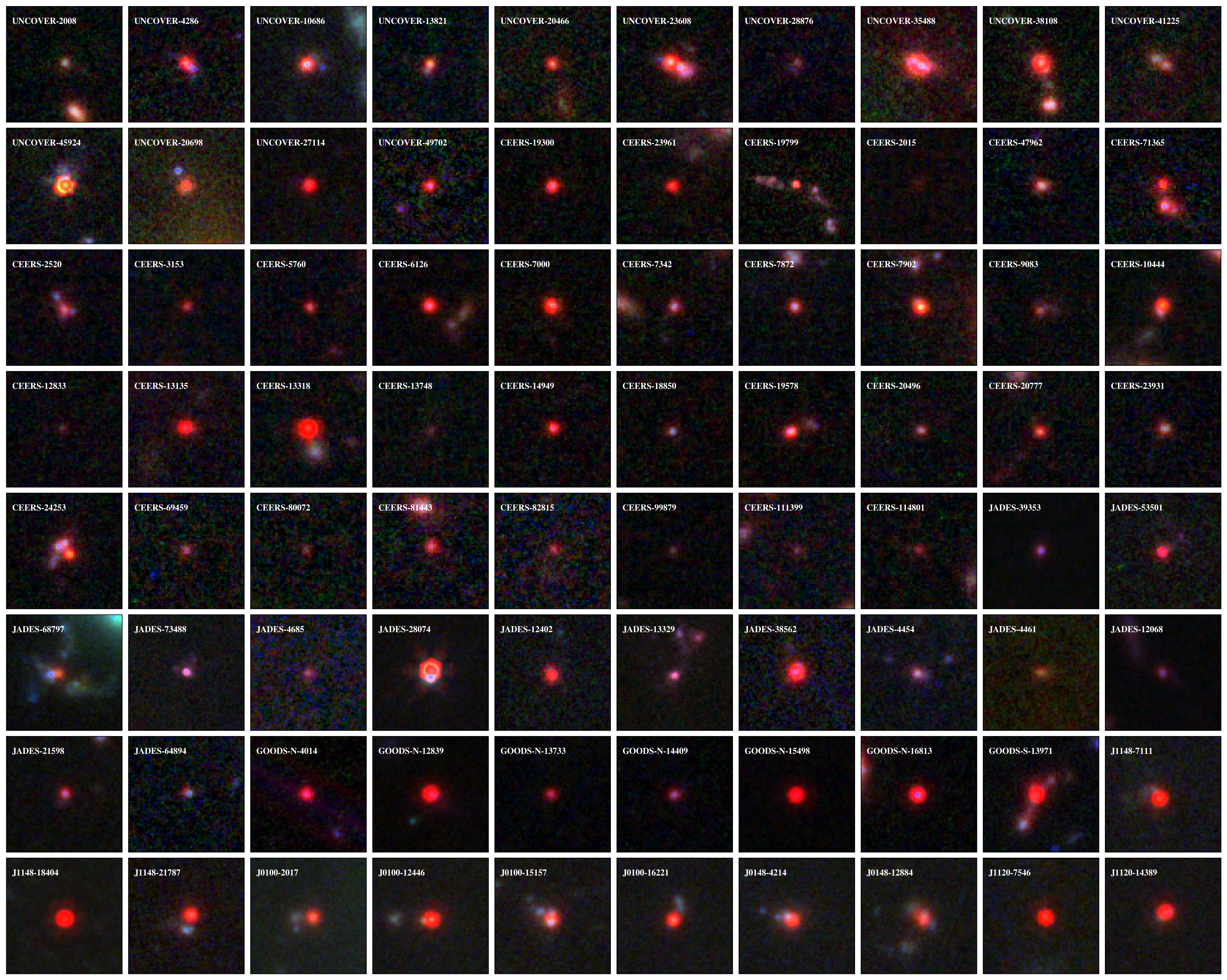}
    \end{minipage}

    \vspace{1.0cm} 

    \begin{minipage}{\linewidth}
        \centering
        {\large\textbf{Strongly Lensed LRDs}}\par
        \vspace{0.2cm}
        \includegraphics[width=\linewidth, trim={0cm 0cm 0cm 0.1cm}, clip]{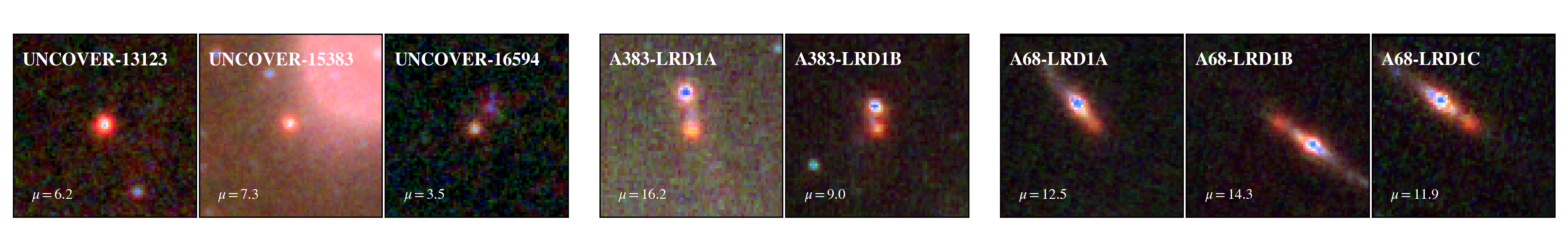}
    \end{minipage}
    \caption{RGB composites of the full LRD sample. All images are $1\arcsec5\times1\arcsec5$, constructed from the available JWST/NIRCam filters for each object, selected as a function of redshift to approximately sample rest-frame UV emission (blue), wavelengths near the Balmer/4000~\AA\ break (green), and redder rest-frame optical emission (red), depending on filter availability.
    For the three strongly lensed LRDs, the lensing magnifications are indicated in the lower left corner.
    The images are shown for illustrative purposes; all quantitative measurements are derived from the structural modeling described in Section~\ref{sec:structuralmodeling}.
    }
     \label{fig:RGB_selection}
\end{figure*}


\begin{figure*}
    \centering
    \begin{minipage}{\linewidth}
        \centering
        {\large\textbf{Literature Sample}}\par
        \vspace{0.2cm}
        \includegraphics[width=\linewidth]{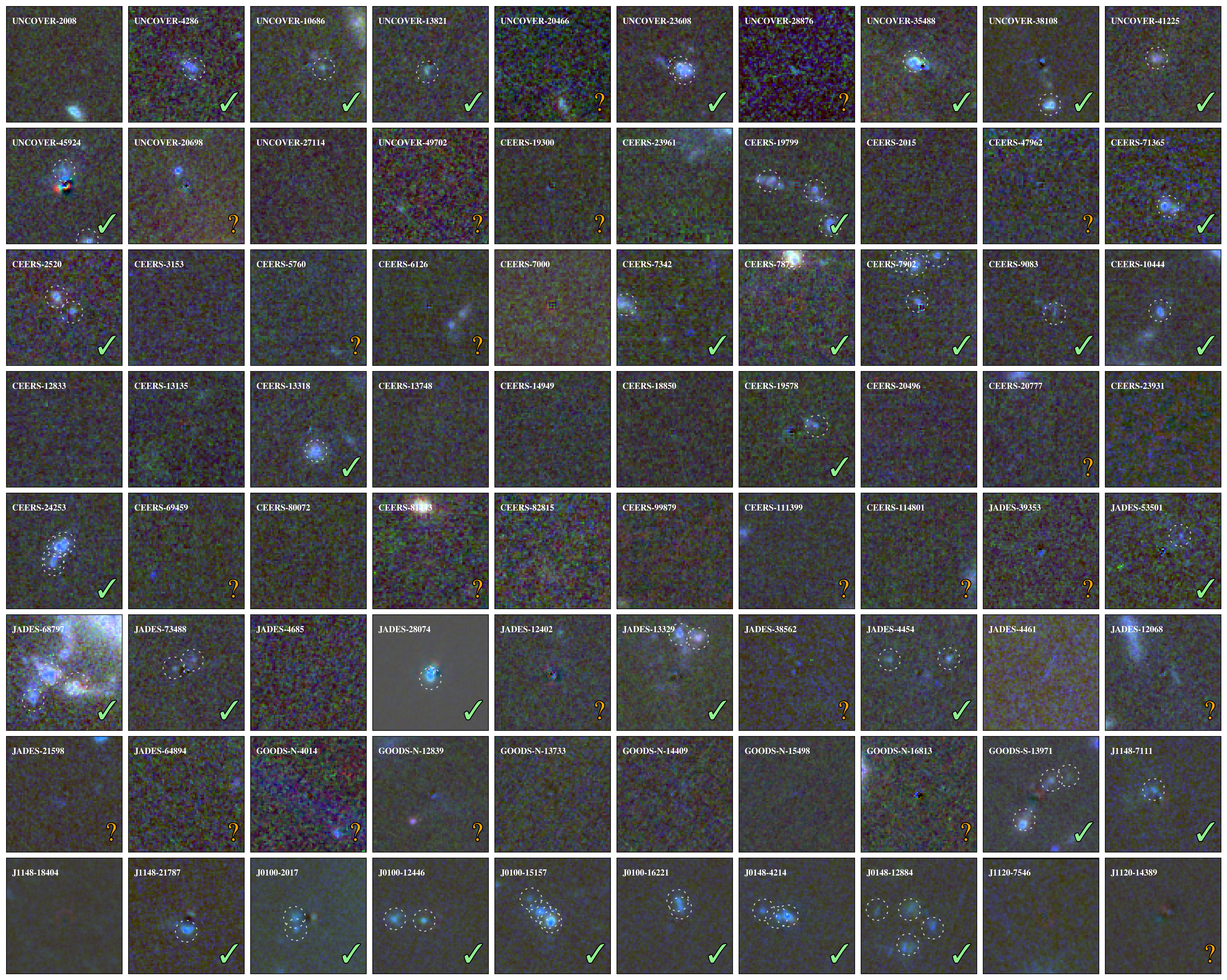}
    \end{minipage}

    \vspace{1.0cm} 

    \begin{minipage}{\linewidth}
        \centering
        {\large\textbf{Strongly Lensed LRDs}}\par
        \vspace{0.2cm}
        \includegraphics[width=\linewidth, trim={0cm 0cm 0cm 0.1cm}, clip]{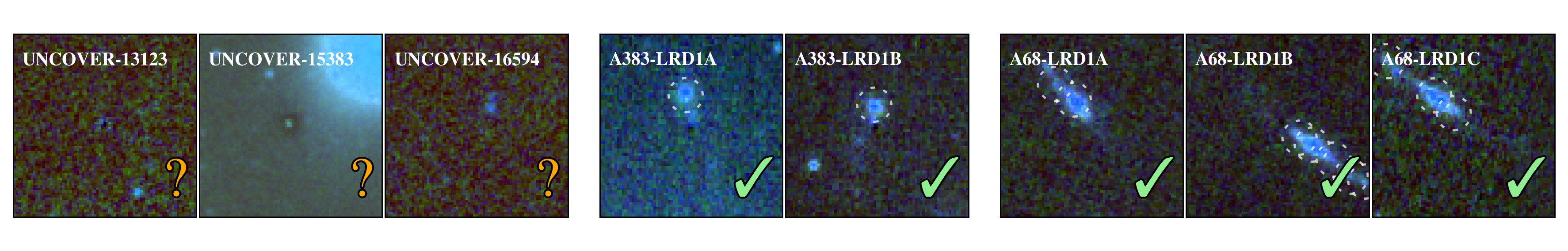}
    \end{minipage}
    \caption{Same as Fig. 1, but showing residual images from the GALFIT modeling after subtracting the red component. The fits are performed in F200W and then applied to the other bands with fixed structural parameters and free magnitudes. To highlight faint UV emission, the images are displayed with a more aggressive scaling and contrast than in Fig. 1. Symbols indicate the companion classification: a check mark ($\checkmark$) denotes a robust companion, while a question mark (?) indicates a tentative candidate, identified either through a dropout consistency or the presence of faint residual emission not robustly isolated as an independent source (see Section~\ref{sec:discussion:companionfraction} for details). The generally low residuals demonstrate that the F200W-based model provides a good representation of the source morphology across bands.
    }
    \label{fig:RGB_selection_blue}
\end{figure*}

\section{LRD Selection and Data}
\label{sec:data:literature}
\subsection{LRDs from Existing JWST Catalogs}
We assemble a comprehensive sample of LRDs by combining catalogs from four major independent studies: \citet{Greene2024}, \citet{Kocevski2025ApJ...986..126K}, \citet{Matthee2024}, and \citet{Graaff2025_unified}. While these catalogs rely on varying selections, the resulting compilation provides a broad and representative cross section of the LRD population as currently defined in the literature.

To ensure reliable physical size measurements, we restrict the sample to sources with secure spectroscopic redshifts. The majority of objects in our compiled sample already have published spectra.
For photometrically selected candidates from \citet{Kocevski2025ApJ...986..126K}, we obtain spectroscopic redshifts by cross-matching with the DAWN JWST Archive (DJA) spectroscopy table (v4.4; using grade 2 or 3)\footnote{https://zenodo.org/records/15472354}. These spectra are reduced using \textsc{msaexp} \citep{Brammer2023}, following the procedures described in \citet{Graaff2024_rubies} and \citet{Heintz2025A&A...693A..60H}.
Duplicate sources between the selections are removed by cross-matching catalogs based on sky position.
To maintain a sample size manageable for detailed structural modeling, we restrict the scope of this work to 
LRDs selected in the Ultradeep NIRSpec and NIRCam Observations before the Epoch of Reionization (UNCOVER) program \citep{Bezanson_uncover2024}, the Cosmic Evolution Early Release Science (CEERS) survey \citep{Finkelstein2025ApJ...983L...4F}, the JWST Advanced Deep Extragalactic Survey (JADES; \citealt{Eisenstein_Jades_2026ApJS..283....6E}) complemented with the First Reionization Epoch Spectroscopically Complete Observations (FRESCO) program \citep[PI: Oesch; PID: GO-1895;][]{Oesch2023}, and the Emission-line galaxies and Intergalactic Gas in the Epoch of Reionization (EIGER) program \citep[PI: Lilly; PID: 1243;][]{Kashino2023}. 


\subsection{Strongly Lensed Little Red Dots}
\label{sec:LRD-A68}
In addition to the literature sample described above, we include two newly discovered strongly lensed, multiply imaged LRDs: A383-LRD1 and A68-LRD1. A383-LRD1 is a striking pair of a red dot and a blue dot, embedded in a low surface brightness structure. It is described in 
\citet{Baggen2025}, \citet{Golubchik2025}, and
\citet{Knudsen2025}. We focus on image A383–LRD1B, which lies in a
cleaner background, less impacted by intracluster light, and enables detailed structural measurements \citep{Baggen2025}, while adopting magnification estimates from \citet{Golubchik2025} based on updated lensing models.

Motivated by the wealth of morphological information that is seen in A383-LRD1 on scales that are typically unresolved, we searched for additional examples of strongly lensed LRDs.  As part of this search we identify a new strongly lensed LRD, triply imaged behind the galaxy cluster A68, similarly observed as part of the VENUS collaboration (Program ID: GO 6882; PI: S. Fujimoto). It is shown in the bottom panels of Fig.\ \ref{fig:RGB_selection}.  
and a larger image of the galaxy cluster with the three images is shown in Appendix~\ref{app:a68_triple} (Fig.~\ref{fig:abell_68_cutouts}).
The system has a secure spectroscopic redshift from the Ly$\alpha$ line (with a luminosity of $\sim3\times10^{42}$ erg/s) of $z=5.421$ \citep{Richard2007} and reported magnifications for the images of $\mu_1=12.5\pm 0.9$, $\mu_2=14.3\pm 0.9$, and $\mu_3=11.9\pm 1.1$. 
It has a similar morphology as A383-LRD1: a close pair of compact objects, one red, one blue. 

While we identified several additional candidate systems with similar morphologies, we exclude them from the analysis as they do not have spectroscopic redshifts. The status of these objects as LRDs is therefore not secure, and their physical sizes and rest-frame luminosities are uncertain. We note that one such candidate has been very recently reported in 
\citet{Yanagisawa2026}, who found a star-forming clump associated with two red dots. 


\subsection{Imaging and Photometry}
All imaging data used in this work are mosaics processed with the \textsc{grizli} pipeline \citep{gabe_brammer_grizli}, version v7.2 or later. This version includes the \texttt{snowblind} routine to mask NIRCam and NIRISS snowballs, as well as an updated bad pixel mask. Most mosaics are publicly accessible via the DJA\footnote{\url{https://dawn-cph.github.io/dja/imaging/v7/}}.
Throughout our analysis, we require short-wavelength mosaics with $0\arcsec02$/pixel sampling and long-wavelength mosaics with $0\arcsec04$/pixel sampling.

For the UNCOVER field, we use the DR3 mosaics available through the UNCOVER data release website \citep{Bezanson_uncover2024,Suess2024_mediumbands}\footnote{\url{https://jwst-uncover.github.io/DR3.html}}. These products are identical to the \textsc{grizli} v7.2 mosaics hosted on DJA, named as \texttt{abell2744clu-grizli-v7.2-{filter}-clear\_drc}.
For the CEERS field, we use short-wavelength mosaics from DJA (v7.2), provided as 12 individual tiles, named \texttt{ceers-full-xi.yj-v7.2-{filter}-clear\_drc}, with $i=0$–5 and $j=0$–1, along with long-wavelength mosaics, \texttt{named ceers-full-grizli-v7.2-{filter}-clear\_drc}. 
The GOODS fields, observed through the JADES and/or FRESCO program, mosaics are similarly drawn from DJA, with filenames \texttt{gds-grizli-v7.2-{filter}-clear\_drc} and \texttt{gdn-grizli-v7.4-{filter}-clear\_drc} for GOODS-S and GOODS-N, respectively.
For the EIGER and VENUS programs, where no public mosaics are currently available, we reduce the images locally using the \textsc{grizli} v7.2 pipeline. These reductions follow the methodology outlined in the PANORAMIC field notebook\footnote{\url{https://github.com/gbrammer/panoramic-jwst/blob/main/Notebooks/step3-panoramic-mosaics.ipynb}}. 

For all LRDs, we adopt published catalog photometry where available. For the UNCOVER field, we use the latest DR3 photometric catalog 
\citep{Bezanson_uncover2024, Suess2024_mediumbands, Weaver2024}, from which we also take the corresponding lensing magnification values.
For CEERS, we use the PR1.1 public catalog\footnote{https://zenodo.org/records/11658282}.
\citep{Weaver2024,Wright2024}. For sources in JADES and/or FRESCO, photometry is taken from the v2.0 and v1.0 public catalogs \citep{Rieke2023}.\footnote{\url{https://archive.stsci.edu/hlsp/jades}; Newer versions have since become available; however, the catalog adopted here was the latest public release at the time the majority of this analysis was performed.}
The photometry for the LRDs detected with the EIGER program is reported in Table 8 in \citet{Matthee2024}; the methodology is reported in \citet{Kashino2023}.

For the newly identified systems presented in this work, we perform our own multiband photometric measurements following the procedure described in \citet{Baggen2025}. In short, we perform point-spread function (PSF)-matched aperture photometry using $0\arcsec5$ circular apertures. All JWST images are convolved to match the F444W PSF, and we apply aperture corrections to both the measured fluxes and uncertainties, which are derived empirically from source-free background regions. 

We emphasize that the photometry is used primarily to measure total fluxes and luminosities of the LRDs, and not as the basis for the component-resolved structural analysis, which relies on direct image modeling. 

Using the imaging products, we exclude 14 LRDs from the parent sample that lack coverage in the key bands required for this analysis. All retained sources therefore have consistent multiband imaging suitable for reliable morphological and photometric measurements.

The final sample consists of 83 LRDs. For consistency, we adopt the source identifiers assigned in the original works from which the LRDs were selected. The sample includes 
three strongly lensed LRDs: the triply imaged LRD in A2744 (UNCOVER-13123, UNCOVER-15383, and UNCOVER-16594), the doubly imaged LRD in A383 (A383-LRD1A and A383-LRD1B), and the triply imaged LRD in A68 (discovered in this work, named A68-LRD1A, A68-LRD1B, A68-LRD1C). Each strongly lensed system is counted only once in the total sample size. For the analysis presented here, we focus on a single representative image for each system, UNCOVER-16594, A383-LRD1B, and A68-LRD1A, to avoid double-counting.
Figure~\ref{fig:RGB_selection} shows red, green, and blue (RGB) image stamps of all sources included in the analysis.

\begin{table*}[t]
\centering
\begin{tabular}{lrrccccccc}
\hline
ID & R.A. & Decl. & $z$ & $\mu$ & $N_{\rm comp}$  & $d_{\rm eff}$ [kpc] & $M_{\rm UV, tot}$ & $M_{\rm LW, tot}$  & $\log_{10}(J_{21})$ \\
\hline
A68-LRD1 & 9.26778 & 9.16196 & 5.42 & 12.5 & 2 & 0.40 & -19.0 & -18.5 & 4.7 \\
A383-LRD1 & 42.01924 & -3.53292 & 6.03 & 9.0 & 1 & 0.34 & -18.1 & -17.5 & 4.4 \\
UNCOVER-4286 & 3.61920 & -30.42327 & 5.84 & 1.6 & 1 & 0.37 & -18.5 & -18.0 & 4.5 \\
UNCOVER-10686 & 3.55084 & -30.40660 & 5.05 & 1.4 & 1 & 0.82 & -17.8 & -17.1 & 3.5 \\
UNCOVER-13821 & 3.62061 & -30.39995 & 6.34 & 1.6 & 1 & 0.44 & -16.2 & -15.3 & 3.3 \\
UNCOVER-23608 & 3.54282 & -30.38065 & 5.80 & 2.1 & 2 & 0.79 & -18.6 & -17.9 & 3.8 \\
UNCOVER-35488 & 3.57898 & -30.36260 & 6.26 & 3.4 & 1 & 0.31 & -18.1 & -16.9 & 4.2 \\
UNCOVER-38108 & 3.53001 & -30.35801 & 4.96 & 1.6 & 1 & 2.80 & -18.4 & -17.9 & 2.7 \\
UNCOVER-41225 & 3.53399 & -30.35331 & 6.76 & 1.5 & 1 & 0.68 & -18.2 & -17.7 & 3.9 \\
UNCOVER-45924$^*$ & 3.58476 & -30.34363 & 4.46 & 1.6 & 2 & 0.61 & -19.1 & -18.6 & 4.3 \\
CEERS-19799 & 214.92415 & 52.84905 & 4.22 & - & 3 & 2.42 & -20.2 & -19.7 & 3.6 \\
CEERS-71365 & 214.89554 & 52.90672 & 4.80 & - & 1 & 1.75 & -19.4 & -19.0 & 3.6 \\
CEERS-2520 & 214.84477 & 52.89210 & 8.69 & - &  2 & 0.74 & -19.5 & -19.0 & 4.3 \\
CEERS-7342 & 215.00849 & 52.97797 & 6.12 & - & 1 & 3.96 & -20.5 & -20.0 & 3.3 \\
CEERS-7872 & 214.87615 & 52.88083 & 8.36 & - & 1 & 3.04 & -20.8 & -20.4 & 3.6 \\
CEERS-7902 & 214.98304 & 52.95601 & 6.99 & - & 4 & 0.87 & -20.2 & -19.7 & 4.5 \\
CEERS-9083 & 214.79753 & 52.81876 & 6.62 & - & 1 & 0.87 & -18.6 & -18.1 & 3.8 \\
CEERS-10444 & 214.89225 & 52.87741 & 6.69 & - & 1 & 0.52 & -19.1 & -18.7 & 4.5 \\
CEERS-13318 & 214.79537 & 52.78885 & 5.28 & - & 1 & 1.89 & -19.9 & -19.4 & 3.7 \\
CEERS-19578 & 214.88016 & 52.81256 & 5.28 & - & 1 & 1.71 & -18.5 & -18.0 & 3.2 \\
CEERS-24253 & 214.97996 & 52.86108 & 6.23 & - & 4 & 1.01 & -20.4 & -19.9 & 4.4 \\
JADES-53501 & 189.29506 & 62.19357 & 3.44 & - & 1 & 2.25 & -17.4 & -16.9 & 2.5 \\
JADES-68797 & 189.22914 & 62.14619 & 5.04 & - & 3 & 0.72 & -20.4 & -19.6 & 4.6 \\
JADES-73488 & 189.19740 & 62.17723 & 4.13 & - & 2 & 1.07 & -17.2 & -16.7 & 3.1 \\
JADES-28074 & 189.06459 & 62.27382 & 2.27 & - & 1 & 0.30 & -19.6 & -18.8 & 5.0 \\
JADES-13329 & 53.13904 & -27.78443 & 3.94 & - & 2 & 3.95 & -19.0 & -18.4 & 2.6 \\
JADES-4454 & 53.16611 & -27.77204 & 6.30 & - & 2 & 2.46 & -17.5 & -17.0 & 2.5 \\
GOODS-S-13971 & 53.13858 & -27.79025 & 5.48 & - & 3 & 2.21 & -19.7 & -19.2 & 3.5 \\
J1148-7111 & 177.10171 & 52.90796 & 4.34 & - & 1 & 0.93 & -19.6 & -19.2 & 4.2 \\
J1148-21787 & 177.02142 & 52.83362 & 4.28 & - & 1 & 1.23 & -19.3 & -18.9 & 3.8 \\
J0100-2017 & 15.05804 & 28.07241 & 4.94 & - & 2 & 1.49 & -17.6 & -17.1 & 3.0 \\
J0100-12446 & 15.04825 & 28.00972 & 4.70 & - & 2 & 1.49 & -19.5 & -18.9 & 3.7 \\
J0100-15157 & 15.03025 & 28.05018 & 4.94 & - & 4 & 0.86 & -20.6 & -20.1 & 4.6 \\
J0100-16221 & 15.03404 & 28.05158 & 4.35 & - & 2 & 1.40 & -19.1 & -18.7 & 3.6 \\
J0148-4214 & 27.13871 & 5.99723 & 5.02 & - & 3 & 0.53 & -20.1 & -19.7 & 4.9 \\
J0148-12884 & 27.17325 & 6.01592 & 4.60 & - & 4 & 1.77 & -20.1 & -19.6 & 3.8 \\
\hline
\end{tabular}
\caption{Physical properties of LRD systems with candidate companions ($N_{\rm comp} \geq 1$).
Listed are the source coordinates, spectroscopic redshift, lensing magnification $\mu$, and the number of companion components identified by our source-finding pipeline within a $1\arcsec5$ cutout (see Section~\ref{sec:componentidentification}).
The effective separation $d_{\rm eff}$ corresponds to the projected distance at which a single companion with the same total LW luminosity would reproduce the combined LW radiation field from all companions (see Section~\ref{sec:lymanwerner}).
$M_{\rm LW,tot}$ denotes the total intrinsic luminosity of the companion population integrated over the LW band (91.2–111 nm), computed from the intrinsic SED templates, while $M_{\rm UV,tot}$ is the corresponding rest-frame 1500$\AA$ absolute magnitude. 
The final column reports the resulting incident LW intensity at the position of the compact red component, expressed as $J_{21,\rm LW}$ (computed from $M_{\rm LW,tot}$ and $d_{\rm eff}$).
$^*$ This source has additional confirmed companions  \citep{Torralba2026}; here we report only the two companions as found within the $1\arcsec5$ cutout. 
}
\label{tab:lrd_sample}
\end{table*}

\section{Structural Modeling and SED Decomposition}
\label{sec:structuralmodeling}
In this section, we describe how we identify distinct structural components associated with each LRD, measure their projected separations, and extract component-resolved SEDs. These measurements form the basis for estimating the local UV radiation field in the following section.

\subsection{Component Identification}
\label{sec:componentidentification}

Using all available imaging, we investigate whether LRDs consist of a single compact source or contain physically distinct components at small projected separations. We developed a source-finding pipeline that identifies such components in two stages.

First, we generate a segmentation mask for the $1\arcsec5$ cutout using standard \texttt{photutils} detection \citep{larry_bradley_2025_14889440} on a high signal-to-noise ratio (SNR) stack of all available bands. This reliably identifies the primary red dot and any clearly separated neighbors. However, the resulting mask for the central LRD is often quite large, potentially obscuring companions  that are distinct in the higher-resolution UV bands and very close but blended at longer wavelengths.

To recover this substructure, we perform a second step by using only the short-wavelength bands (F070W-F150W). We make a new segmentation mask, allowing lower SNR $>3$, and separated from the red centroid by at least 1 pixel. 
This approach allows us to robustly identify single compact companions, multiple knots, or faint extended emission next to the compact red dot that would otherwise be lost within the central mask. 

Once components are identified, we assess their physical association with the LRD. For blue knots recovered directly within the central mask (via the second step), the projected separation is extremely small. We therefore assume these are intrinsic parts of the LRD system. 
This interpretation is further supported by the frequent presence of extended emission aligned with the same direction in the full multiband images, as well as by visual confirmation that these components exhibit consistent dropout signatures, with vanishing flux shortward of the Lyman limit.

For more distant neighbors outside the central mask, we rely on photometric redshifts derived using \texttt{EAZY} \citep{Brammer2008} to rule out chance projections. We first perform unconstrained free-redshift fits. 
For sources with broad filter coverage, these fits often converge robustly. For example the outer knot in the bottom of UNCOVER-45924 converges to $z=4.46$. Notably, this companion has since been spectroscopically confirmed by \citet{Torralba2026}, who also identify additional companions at larger projected separations, lending confidence to both our source detection and photometric redshift methodology.
However, for fainter sources or those lacking strong dropout constraints, the free-redshift solutions can be poorly constrained. In such cases, we perform a second fit with the redshift fixed to the LRD spectroscopic redshift, $z_{\rm spec}$. A neighbor is classified as a companion if the fixed-redshift fit is statistically consistent with the data, defined by $\chi^2(z_{\rm spec}) - \chi^2 (z_{\rm free}) < 1$. For these systems, the physical association remains less secure and should be interpreted with caution until spectroscopic confirmation becomes available.

For the triply imaged system in A2744, the interpretation of companions requires additional care. While companions are identified for the images UNCOVER-13123 and UNCOVER-16594, no companion is detected for UNCOVER-15383, and the properties of the candidate companions differ between the images. We therefore treat these companions as tentative. Further confirmation will require a detailed delensing analysis to the source plane and consistency checks across all multiple images.

In total, 36 LRDs ($43\%$ of the sample) show evidence for at least one physically associated companion ($N_{\rm comp} \geq 1$). 
For these systems, the identified companions are encircled in Fig.~\ref{fig:RGB_selection_blue}, which shows residual images from the GALFIT modeling after subtraction of the red component (see the following section for details).
This component identification step defines the set of UV sources used to estimate the local LW radiation fields, which we quantify in the following sections. Systems with $N_{\rm comp} \geq 1$ are summarized in Table~\ref{tab:lrd_sample}, while LRDs without identified companions are listed in Appendix \ref{app:sources_no_comp} (Table \ref{tab:appendix_table}). 

\begin{figure*}
    \centering
     \raisebox{0.03\textheight}{%
     \hspace{0.5cm}
      \includegraphics[width=0.64\linewidth]{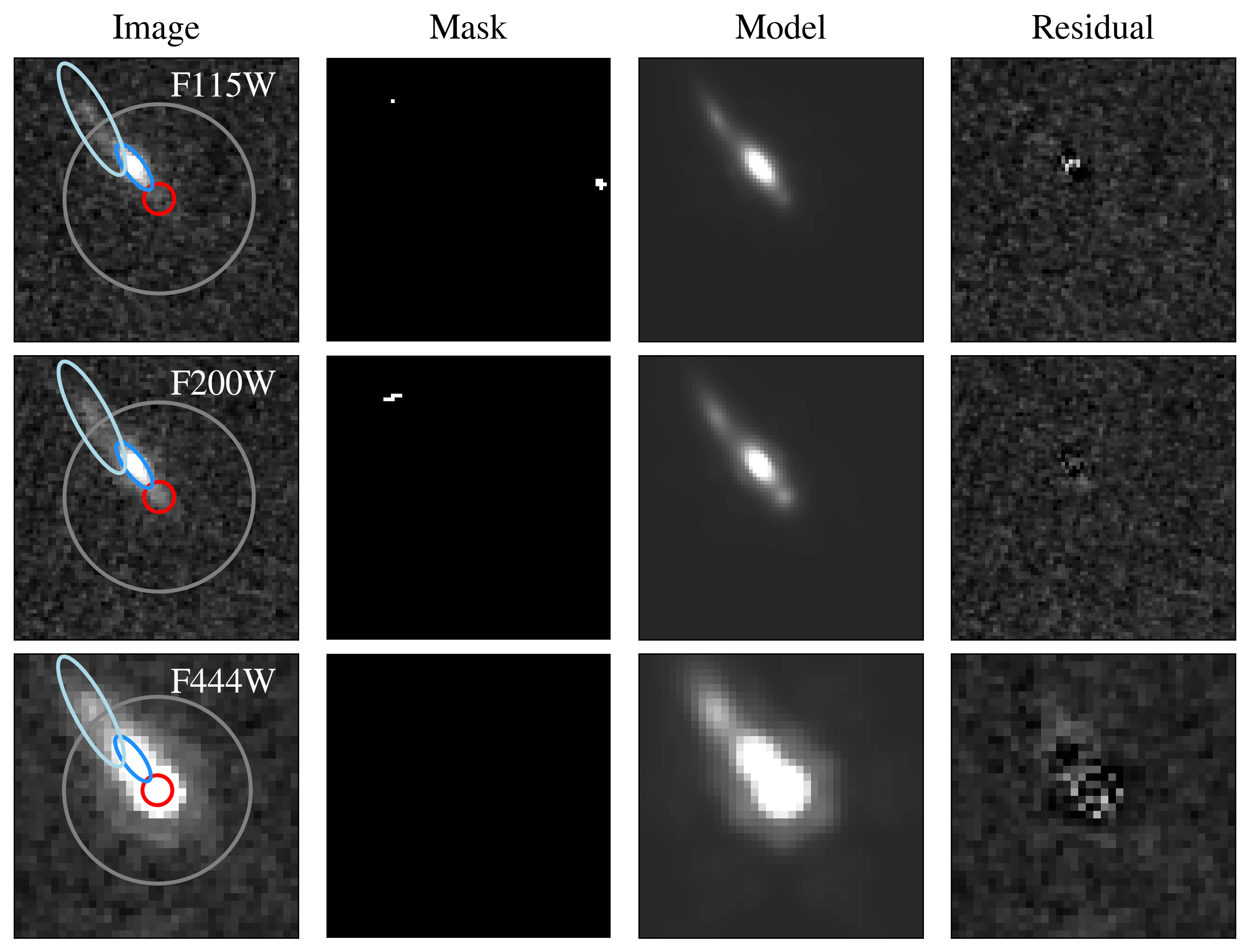}
    }  
    \includegraphics[width=0.48\linewidth]{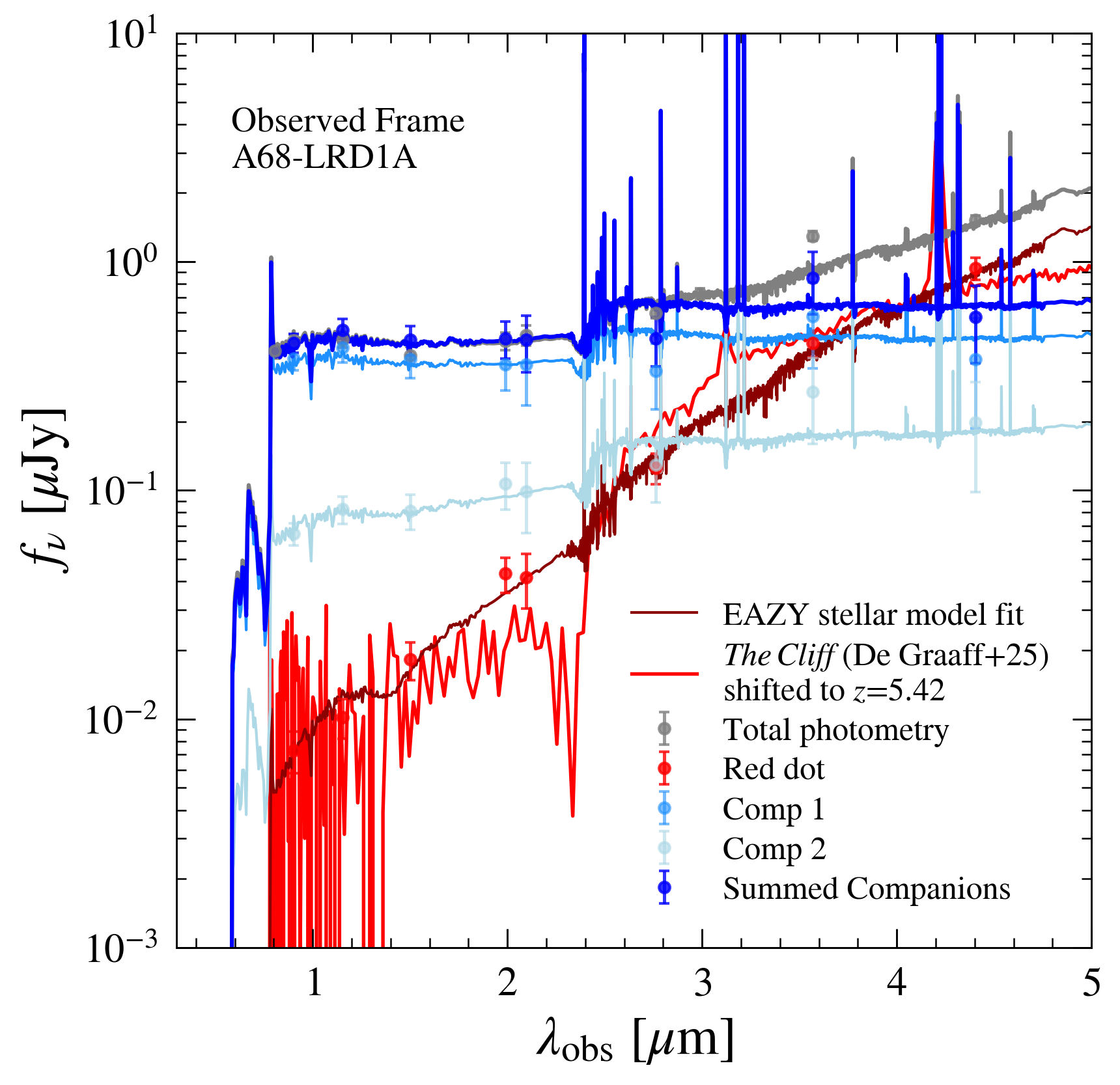}
    \includegraphics[width=0.48\linewidth]{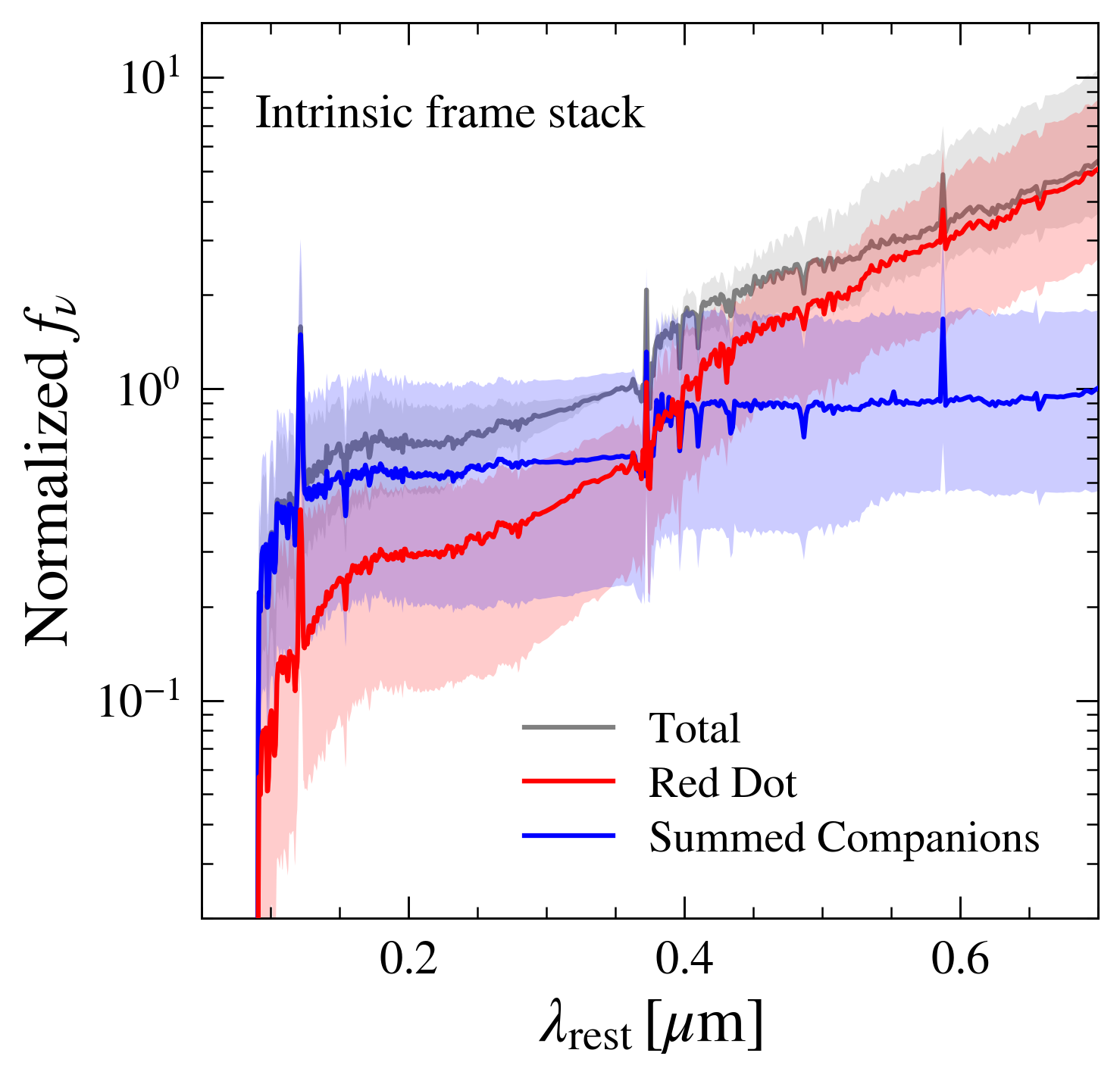}
    \caption{ \textit{Top:} JWST/NIRCam cutouts in three filters (F115W, F200W, F444W), the corresponding segmentation mask, best-fit GALFIT model, and residual image for A68-LRD1A. Ellipses indicate the fitted component parameters from F200W, overlaid on all bands. The gray circle shows the $0\arcsec5$ photometric aperture. \textit{Bottom left:} SED decomposition for A68-LRD1A. Gray points show the observed total aperture photometry, while the red and blue points show the forced photometry obtained from the GALFIT fits shown in the top panel, adopting the same color scheme. Solid curves show the best-fit EAZY templates. These SEDs are shown in the observed frame, meaning that they include lensing magnification and intergalatic medium absorption.
    Overplotted is \textit{The Cliff} spectrum from \citet{Graaff2025_Cliff}, only corrected for the redshift difference (without any renormalization), showing remarkable resemblance to the SED shape \textit{and} intrinsic luminosity. 
    \textit{Bottom right:} stacked intrinsic rest-frame SEDs for all LRDs with companions. For each system, the total SED (gray) is normalized to the total flux density at a rest-frame wavelength of 0.3645$\mu$m (Balmer break). The red and blue component SEDs are shown relative to this normalization. Shaded regions indicate object-to-object scatter (16th–84th percentile). For systems with multiple companions, the blue curve represents the sum of all companions. 
    }
\label{fig:decomposition_example}
\end{figure*}
\subsection{S\'ersic Profile Fitting}
\label{sec:sersic_profile_fitting}
To measure the projected separations between the compact red component and any associated companions, we model the surface brightness distributions of the systems using two-dimensional S\'ersic profile fitting with GALFIT \citep{Peng2002, Peng2010x}. Structural modeling is performed in the F200W band, which provides the highest SNR at optimal angular resolution. For a small number of sources
without F200W coverage, we instead use the F210M image.

We model each system using a multicomponent S\'ersic fit consisting of the central red dot and its associated companion(s), as identified by our initial source-finding pipeline. The number of fitted components is therefore $N_{\rm fit} = N_{\rm comp} + 1$, where $N_{\rm comp}$ the number of companions and the additional component corresponds to the red dot itself. For each component, we optimize the centroid position $(x, y)$, effective radius ($r_{\rm e}$), integrated magnitude, axis ratio ($b/a$), and position angle (PA). Initial centroid positions are taken from the pipeline but are allowed to vary freely during the fitting. To ensure stability in the multicomponent fits, we fix the S\'ersic index to $n = 1.5$ for all components. Parameter ranges are restricted to $r_{\rm e} = 0.5$–$50$ pixels, $b/a = 0.1$–$1.0$, and magnitudes between 1 and 100. 
For the compact red component, we further restrict the effective radius to $r_{\rm e} \leq 10$ pixels, given its compact nature. For the strongly lensed, triply imaged system A68-LRD1, we impose additional constraints to avoid unphysical solutions driven by lensing distortions. In this case, the centroid position of the red component is fixed during the fit and set to the position of the red emission identified by the source-finding pipeline. We also require $b/a > 0.5$ to prevent the model from converging to highly elongated profiles. These constraints are motivated by the observed morphology of the red component, which appears compact and spheroidal in the long-wavelength NIRCam bands.
Prior to fitting, we use sigma-clipped background statistics to identify and mask contaminating foreground or background sources. From this mask, we then ensure that any potential companion is excluded.

For the lensed systems (UNCOVER sources, A383-LRD1, and A68-LRD1) we apply magnification corrections to all measured physical quantities. The adopted magnification values ($\mu$) are listed in Table~\ref{tab:lrd_sample}. 
For projected separations between components, we divide the observed distances by $\sqrt{\mu}$. However, in several strongly lensed cases, most notably A68-LRD1, the separation direction lies approximately along the arc and may therefore be closer to the total magnification. In such cases, the true separations could be smaller than assumed, and the corresponding radiation field (which scales as $1/d^2$) would be correspondingly stronger.
Observed fluxes are divided by $\mu$.

This procedure yields a consistent set of component centroids and projected separations for all systems, which we use in the analysis that follows.


\subsection{Forced Photometry and Component SEDs}
\label{sec:forcedphotometry}
Having identified the relevant components and measured their projected separations, we next determine the UV emission associated with the compact red source and any nearby companions.

Using the fit in F200W/F210M, we then perform forced photometry on the remaining bands to disentangle the SEDs of the components, following the methodology of \citet{Baggen2025}. We fix all structural parameters ($x,\, y,\, r_{\rm e},\, n,\, b/a,\, \rm PA$) to the values determined from the initial fit and allow only the magnitude to vary. This ensures that flux is extracted from consistent physical regions across all filters, allowing us to robustly "split" the SEDs of the red dot and offset emission. Ideally, such decomposition would be achieved through spatially resolved spectroscopy. In the absence of such data, forced multiband photometry provides a valuable tool to isolate the light of the individual components using the available imaging.

Figure~\ref{fig:decomposition_example} illustrates this procedure for A68-LRD1A. The left panel shows the F115W, F200W, and F444W cutouts alongside the segmentation mask, model, and residuals. The ellipses indicate the (fixed) positions, sizes, and orientations of the fitted components derived from F200W and overlaid on all bands. The gray circle is shown to highlight the $0\arcsec5$ aperture used to measure the total photometry. The forced photometry produces remarkably stable fits across all filters despite the fixed geometry.

The middle panel shows the observed-frame SED decomposition for the same system. The gray points and corresponding solid curve represent the total aperture photometry and the associated best-fit EAZY template. To isolate the distinct physical contributions within the system, we plot the forced photometry for the individual components as colored points, with the solid curves representing the corresponding best-fit EAZY templates.  The red component exhibits an extremely steep red spectrum, closely resembling the spectral shape reported for \textit{The Cliff}-LRD \citep[][overplotted]{Graaff2025_Cliff}. This spectrum is corrected only for the redshift difference, without any renormalization, and therefore matches not only the spectral shape but also the intrinsic luminosity.
While EAZY interprets the red component as a smooth continuum from a stellar population and does not capture the extreme Balmer absorption, the close resemblance nonetheless highlights the diagnostic power of strong lensing, which allows the light from the system to be spatially separated and analyzed in detail, even using photometry alone.


This decomposition enables us to quantify the relative contributions of the compact red source and the nearby companions across wavelength. We perform this SED decomposition for all systems, shift each to the rest frame, and normalize by the total SED flux at the Balmer break (rest-frame 0.3645~$\mu$m), as shown in the right panel. We interpolate each component's template onto a common logarithmic rest-frame wavelength grid spanning 0.05–0.7~$\mu$m with 800 points. The shaded regions present the 16th–84th percentile range computed at each wavelength across the sample. The solid curves show the median of the distribution. For LRDs with more than one companion, the blue curve represents the sum. The stacked rest-frame SEDs reveal a consistent picture: the red components dominate at $\lambda_{\rm rest} \gtrsim 0.4$~$\mu$m with steeply rising continua, while the companions contribute primarily at UV wavelengths ($\lambda_{\rm rest} \lesssim 0.3$~$\mu$m) with relatively flat spectra. 
The companions are therefore best interpreted as young, star-forming systems with modest stellar masses (typically of order $\sim10^{8}$–$10^{9}\msun$, with a mean for the total sample of $\langle \log (M_\star/\msun)\rangle = 8.8$).
Some individual cases can reach higher inferred stellar masses, such as UNCOVER-45924 ("Monster" LRD), with a stellar mass of $\sim10^{10.2}\msun$, despite separating the UV emission from the red component. This remains lower than the mass inferred when the full SED, including the Balmer break, is fit \citep[$\sim10^{10.9}\msun$][]{Labbe2024}. Another massive example is JADES-68797, with $M_*\sim10^{10.5}\msun$.

We note that the stacked red component SED is generally less steep than the individual strongly lensed cases (which are more comparable to \textit{The Cliff}). This likely reflects residual blending with surrounding stellar light, from either the host galaxy or nearby companions (see Discussion Section \ref{sec:discussion_concordancepicture}), which is difficult to fully remove in unlensed systems.

Together, the SEDs from the companions and projected separations derived here provide the observational inputs required to estimate the local LW radiation field experienced by the compact red components, which we quantify in the next section.

\begin{figure*}[t]
    \centering
    \hspace{1cm}
    \includegraphics[width=0.85\linewidth]{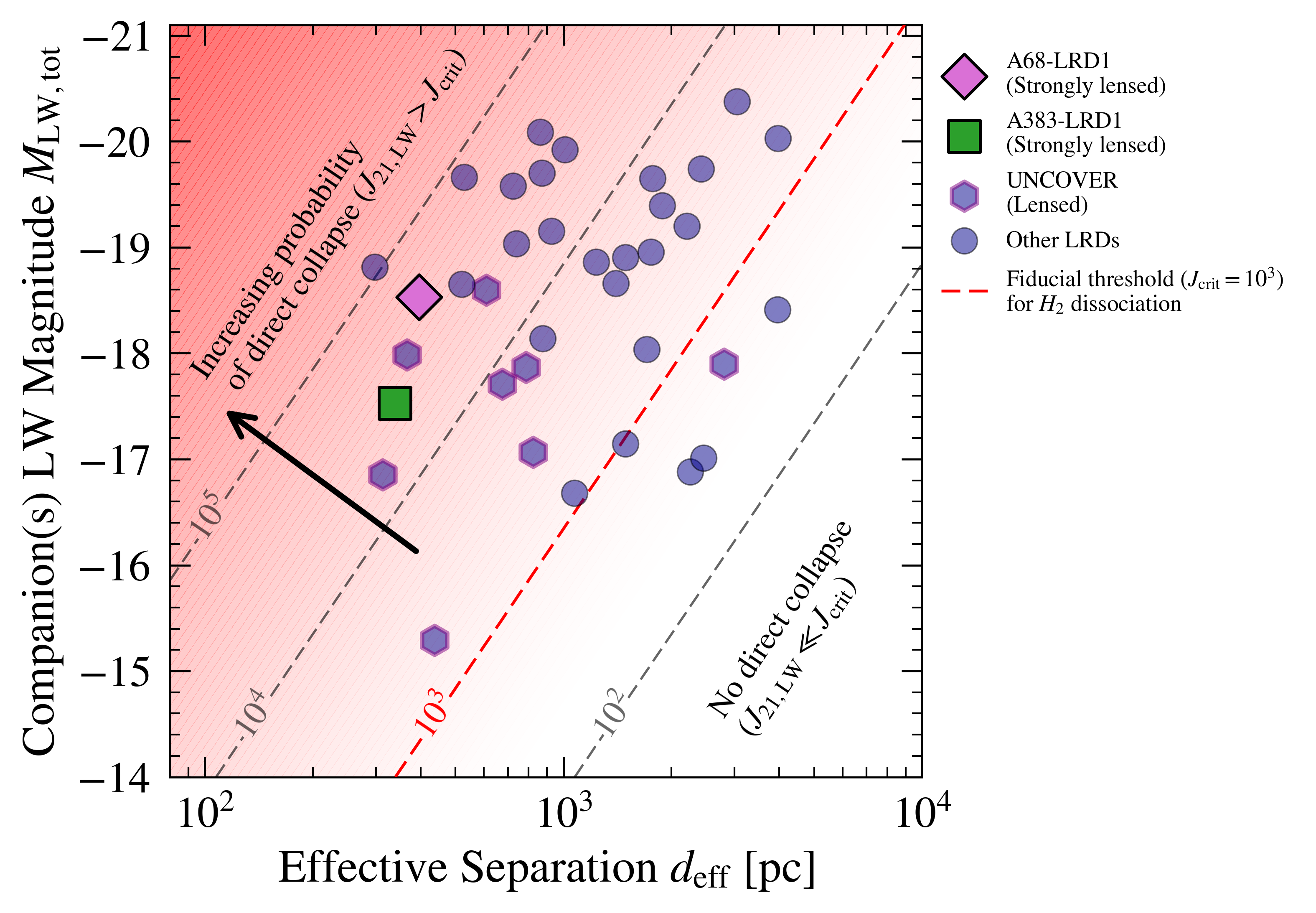}
    \caption{The total LW magnitude of the companion(s), $M_{\rm LW,tot}$, is shown as a function of the effective projected separation from the compact red component, $d_{\rm eff}$ (defined such that $L_{\rm LW,tot}/d_{\rm eff}^2 = \sum_i L_{{\rm LW},i}/d_i^2$ when an LRD has multiple companions).
    Diagonal lines indicate constant LW radiation intensity, $J_{21,\rm LW}$, incident on the red component.
    Systems toward the upper left correspond to brighter companions at smaller separations and therefore higher LW intensities.
    The red dashed line marks a commonly adopted critical threshold for molecular hydrogen dissociation ($J_{\rm crit}=10^3$), above which fragmentation is expected to be strongly suppressed.
    Shaded regions illustrate increasing likelihood for collapse in an LW-regulated regime.
    }
    \label{fig:J21}
\end{figure*}

\section{Lyman-Werner Radiation from the Companions}
\label{sec:lymanwerner}
We now estimate the LW radiation field at the position of the red component. 
For each companion $i$, we take the best-fit SED template obtained through the forced multiband photometric decomposition (Section~\ref{sec:forcedphotometry}). In the EAZY fitting, we disable intergalactic medium absorption, such that the templates represent the emitted UV spectrum rather than the attenuated observed flux. From this, we compute the mean flux density in the rest-frame LW band, defined as $91.2$--$111\,\mathrm{nm}$, and convert it to an intrinsic LW luminosity, using the luminosity distance at the source redshift:
\begin{equation}
L_{\nu,i}^{\rm LW} = \frac{4\pi d_L^2}{1+z}
\left\langle f_{\nu,i} \right\rangle_{\rm LW}.
\end{equation}

By assuming isotropic emission from each companion and dilution over the projected separation ($d_{i}$), we obtain the LW radiation field at the position of the red component,
\begin{equation}
J_\nu = \frac{1}{4\pi}
\sum_i \frac{L_{\nu,i}^{\rm LW}}{4\pi d_{i}^2},
\end{equation}
which is often expressed in dimensionless form as
\begin{equation}
J_{21,\rm LW} \equiv
\frac{J_\nu}{10^{-21}\,
\mathrm{erg\,s^{-1}\,cm^{-2}\,Hz^{-1}\,sr^{-1}}}.
\end{equation}

For convenience, we define the total LW specific luminosity of all companions (if more than one) as
\begin{equation}
L_{\nu,\rm LW}^{\rm tot} = \sum_i L_{\nu,i}^{\rm LW},
\end{equation}
and an effective separation $d_{\rm eff}$ such that
\begin{equation}
\frac{L_{\nu,\rm LW}^{\rm tot}}{d_{\rm eff}^2}
=
\sum_i \frac{L_{\nu,i}^{\rm LW}}{d_i^2}.
\end{equation}
By construction, the LW radiation field computed from
$(L_{\nu,\rm LW}^{\rm tot}, d_{\rm eff})$
is identical to that obtained by summing the contributions of individual companions. This reparameterization allows us to compare LRDs that have different numbers of companions in a consistent way.

For comparison with more commonly used UV diagnostics, we also measure the total intrinsic rest-frame UV magnitude at 1500$\AA$, $M_{\rm UV,tot}$, computed from the same intrinsic SED templates. All derived quantities,  including projected separations, $M_{\rm UV,tot}$, $M_{\rm LW,tot}$, and
the inferred $J_{21,\rm LW}$, are reported in Table~\ref{tab:lrd_sample}.

Figure~\ref{fig:J21} shows the companion LW magnitude ($M_{\rm LW,tot}$) as a function of separation $d_{\rm eff}$. The diagonal contours indicate constant $J_{21,\mathrm{LW}}$, illustrating that sources of different luminosities can generate comparable radiation fields depending on their distance to the red component.

Across the sample, we infer LW radiation fields of $J_{21,\mathrm{LW}} \sim 10^{2.5}-10^5$. Such intensities far exceed the global UV background, which is expected to be of order $J_{21}\sim1$ at $z\sim10$ \citep[e.g.,][]{Haiman_2000,GreifBromm2006MNRAS.373..128G}, and in most cases exceed the commonly adopted threshold for fragmentation suppression ($J_{\rm crit} \sim 10^{3}$). This places the red components in a regime where molecular hydrogen cooling is expected to be strongly suppressed, allowing the gas to approach the atomic-cooling limit without vigorous fragmentation and star formation \citep{Haiman_1997,Dijkstra2008, Shang2010, Dijkstra_2014}.

The above estimates for the LW radiation field do not account for self-shielding effects \citep{Wolcott-Green2017} due to a column of gas between the blue and red components. We test whether this affects our conclusions by estimating the timescale for molecular hydrogen dissociation in this column.
For a red component exposed to an intrinsic LW radiation field $J_{\rm 21,LW}$ produced by its companion(s), the corresponding LW photon production rate of the companions can be written as $\dot{N}_{\rm LW} {=} 3.5 \times 10^{48}(4\pi)(d_{\rm eff}/\rm kpc)^2 J_{\rm 21,LW}$ photons s$^{-1}$.
Assuming the LW photons dissociate $\rm H_2$ with 10\% efficiency, we can estimate the propagation of the H$_2$ dissociation front through a sphere with radius $d_{\rm eff}$, following equation (2) of \citet{Sullivan_2025}, as
\begin{equation}
    4\pi R^2 n_{\rm H_2} dR = (0.1 \dot{N}_{\rm LW} - 4 \pi k_9 \int n_{\rm H} n_{\rm e} r^2 dr) dt,
\end{equation}
with $k_9$, the H$_{2}$ formation rate, given in Table (A1) of \citet{Oh2002}.  Following the model used in \citet{Scoggins_2024}, we estimate $n_{\rm H}$ assuming that the primordial gas is initially compressing adiabatically, giving a maximum central number density $n_{\rm c} \sim 6 (T_{\rm vir}/1000 \rm K)^{3/2}$ cm$^{-3}$ \citep{Visbal2014} and residual electron fraction $n_{\rm e}/n_{\rm H} = 1.2  \times 10^{-5} \sqrt{\Omega_{\rm}}/(\Omega_{\rm b} h)$ \citep{Peebles_1993}. We approximate the H$_2$ abundance using equation (3) of \citet{Scoggins_2024}. We assume that the gas that hosts the red component has not begun fragmentation and is approaching the atomic-cooling limit, with $T{\sim} 8000$K, and is experiencing a faint background LW radiation $J_{\rm21,LW} = 0.1 {\rm J_{21}}$. 

Once the blue companion turns on, the gas out to the red component is irradiated by an LW field $J_{21, \rm LW} \gtrsim$J$_{\rm crit}$.  We find that for a representative separation of $d_{\rm eff} = 1$ kpc, molecular hydrogen is fully dissociated within ${\sim} 10,000$ yr.  We conclude that self-shielding effects likely become negligible by the time the red clump begins its collapse, since the dynamical timescales are much longer than this ${\rm H_2}$ dissociation timescale.

\begin{figure*}
    \centering
    \includegraphics[width=\linewidth, trim ={2cm 1cm 2cm 4cm}, clip]{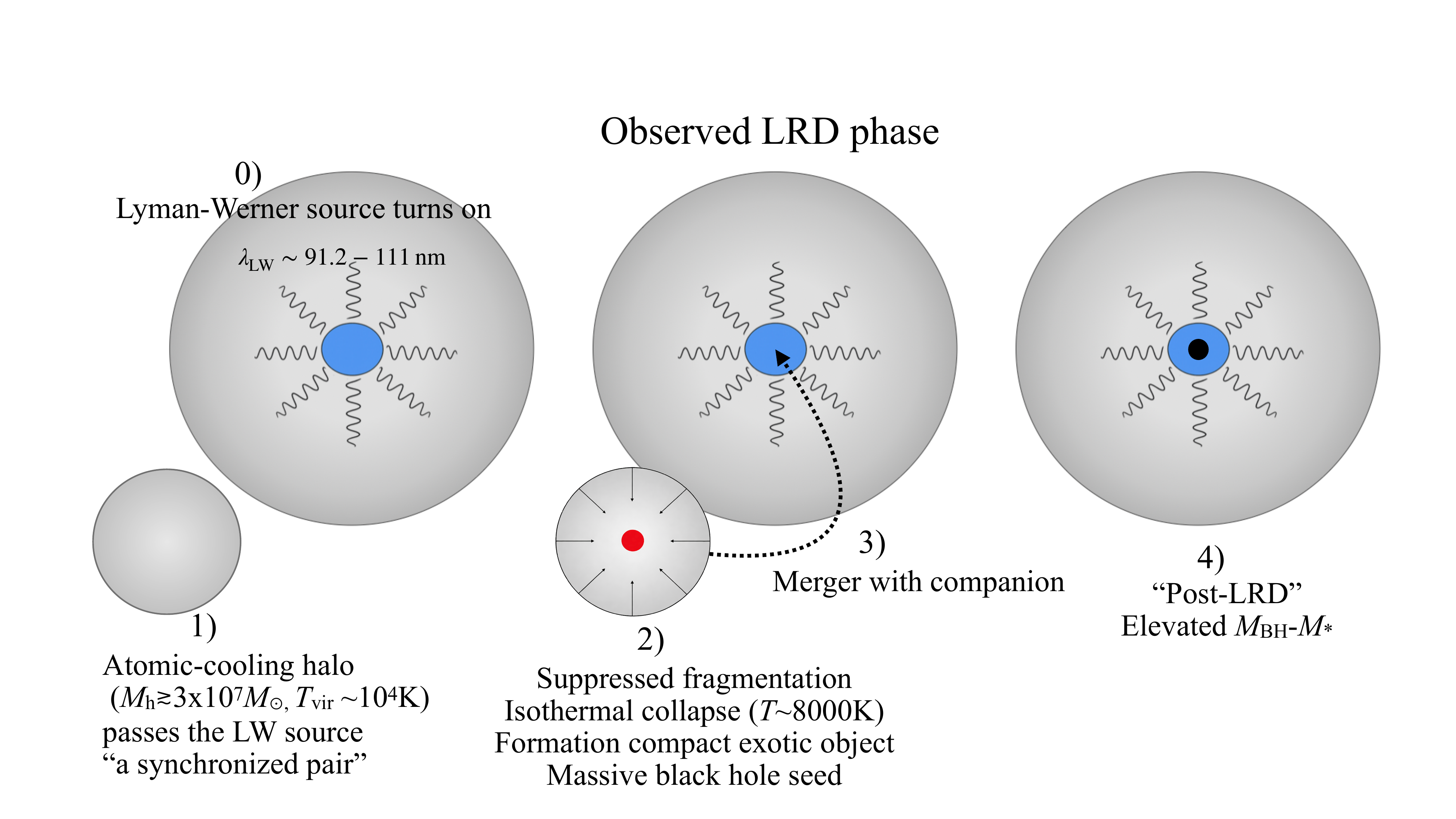}
    \caption{Cartoon illustrating a plausible formation and evolutionary pathway for LRDs. A nearby LW radiation source (0) suppresses gas fragmentation in a neighboring atomic-cooling halo (1), enabling its rapid collapse. Under such conditions, the collapsing gas can form an exotic compact object (2) through phases such as a supermassive star or quasi-star, ultimately growing into a massive black hole seed of order $10^{5}-10^{6}\,\msun$ (i.e., a DCBH; see references in the Sections~\ref{sec:introduction} and \ref{sec:discussion}), observed as an LRD. 
    The compact red source thus forms offset from a UV-bright companion that provides the LW radiation field (2), likely in a lower-mass satellite halo (see Section~\ref{sec:discussion}), producing a  configuration consistent with the morphologies observed in $\gtrsim$43\% of LRDs.
    As the system evolves, the halos may merge (3), while the system may continue to be identified as an LRD as long as it retains a compact, red appearance, before transitioning into a more typical galaxy-AGN system (4).  
    This pathway may represent a key channel for the formation of LRDs, as well as the seeds of the supermassive black holes observed in the local Universe. 
    }
    \label{fig:cartoon}
\end{figure*}

\section{Discussion}
\label{sec:discussion}
\subsection{A Concordance Picture for LRDs: Direct Collapse Induced by Intense Lyman-Werner Radiation}
\label{sec:discussion_concordancepicture}
A substantial fraction of LRDs (43\%) are found in close proximity to one or more UV-bright companions. We show that these companions generate intense local Lyman--Werner radiation fields at the positions of the compact red components, with inferred intensities of $J_{21,\mathrm{LW}} \sim 10^{2.5}$--$10^{5}$. These values are comparable to, and often exceed, thresholds required to suppress molecular hydrogen cooling and fragmentation.




The combination of intense local radiation fields and the frequent occurrence of close companions (see also references in the Introduction) strongly suggests that this is not incidental, but instead reflects a necessary environmental condition for the formation of LRDs.

In this picture, intense local LW radiation suppresses molecular cooling in a nearby atomic-cooling halo (Figure~\ref{fig:cartoon}; step 1), preventing fragmentation and enabling rapid collapse to high densities. 
The immediate products of such collapse are expected to be supermassive stars that may subsequently evolve into quasi-stars once a central black hole forms through core collapse, often referred to as ``direct-collapse" scenarios (Figure~\ref{fig:cartoon}; step 2).
If fragmentation is delayed but not entirely suppressed, 
the same collapse process can also yield extremely dense stellar systems, providing another potential pathway toward massive black hole formation (see references in the Introduction).
The relative importance of these outcomes depends on metal enrichment, gas inflow rates, angular momentum transport, and small-scale fragmentation physics \citep[e.g.,][]{Omukai2008, DevecchiVolonteri2009ApJ...694..302D,
Regan2009MNRAS.396..343R,
Volonteri2010_review,
Ferrara2014MNRAS.443.2410F,
Inayoshi2014MNRAS.445.1549I,Latif2016ApJ...823...40L,Chon2020MNRAS.494.2851C,Chon2025MNRAS.539.2561C}.

A key geometric prediction of this scenario is that the collapse occurs in a nearby (likely a satellite; see \citealt{Visbal2014}) halo, offset from the star-forming galaxy that generates the Lyman–Werner radiation field. The resulting compact object is therefore born spatially offset, naturally producing a configuration in which a compact red component lies close to, but not coincident with, a UV-bright companion, as illustrated schematically in Figure~\ref{fig:cartoon} (step 2).

This framework naturally produces a massive (up to $10^6~{\rm M_\odot}$) black hole, still surrounded by its extremely dense natal cloud, which connects several otherwise puzzling observational properties of LRDs. The extreme compactness of the red components follows directly from collapse in an LW-regulated regime, in which gas approaches the atomic-cooling limit without fragmenting into a normal stellar population. The collapse can also naturally trigger a gas compaction event \citep{Cenci2025}, creating a dense, optically thick environment around the central accreting object. Such conditions provide a plausible explanation for the observed X-ray weakness \citep[e.g.,][]{Ananna2024, MadauHaardt2024ApJ...976L..24M,PacucciNarayan2024ApJ...976...96P,Yue2024,Maiolino2025MNRAS.538.1921M_chandra}, as well as the extremely steep, in some cases nearly blackbody-like, rest-frame optical continua with Balmer breaks, Balmer absorption, and exponential wings observed in many systems. These spectra have been attributed to accreting black holes embedded in dense gas configurations, variously described as cocoons or envelopes (sometimes phenomenologically dubbed BH*); however, the exact geometry of this dense gas structure remains debated 
\citep{Juodzbalis2024_rosetta,
Graaff2025_Cliff, 
Graaff2025_unified, 
DEugenio2025,
InayoshiMaiolino2025,
Ji2025,
Naidu2025_BHstar,
Kido2025MNRAS.544.3407K,
Liu2025ApJ...994..113L, Liu2026arXiv260302317L, MadauMaiolino2026arXiv260222386M, Matthee2026arXiv260317667M, Rusakov2026Natur.649..574R, Sneppen2026arXiv260118864S, Wang2026_water}. 
In particular, the DCBH model of \citet{Pacucci2026arXiv260114368P} demonstrates that accretion through a dense, compressionally heated, collisionally ionized flow can self-consistently reproduce these spectral properties. Similar spectral shapes are also obtained in MESA-based stellar evolution models of quasi-stars \citep{Santarelli2026ApJ...998L...4S_LRD} and in SED models of supermassive stars \citep[e.g.,][]{Chisholm2026arXiv260215935C,NandalLoeb2026ApJ...998..124N}.

In addition, it provides a natural explanation for the characteristic V-shaped SEDs seen in the integrated light of many LRDs. In this picture, the V-shape arises from the superposition of the compact red component and UV-bright host \citep[see also][]{Chisholm2026arXiv260215935C}, as directly demonstrated by strongly lensed, spatially resolved systems \citep{Baggen2025} (see also Figure~\ref{fig:decomposition_example}). Without the aid of strong lensing, such configurations generally remain unresolved at JWST resolution, causing the combined emission to appear as a single compact source with distinctive colors. In hindsight, the commonly used photometric selection of LRDs has proven highly effective at identifying candidate collapse systems, as it implicitly selects objects in which compact red components are in very close proximity to UV-bright companions, thereby explaining why the joint requirement of compactness and V-shaped colors always selects broad-line sources \citep{Hviding2025}.

From a complementary perspective, \citet{Sun2026arXiv260120929S} interpret LRDs as systems consisting of BH*'s embedded within a surrounding host galaxy, based on fits to their integrated spectral emission. Our results provide a morphological basis for this interpretation by demonstrating that the red and blue components are often spatially distinct and that the steep red continuum (BH*-like; Figure~\ref{fig:decomposition_example}) can be isolated from surrounding star-forming emission. From our component-resolved SED fitting, we find that the companions typically have modest stellar masses of order $M_\star \sim 10^{8}$–$10^{9}\msun$, consistent with the stellar masses inferred from narrow emission lines \citep[e.g.,][]{Deugenio2025_lrd_furtak,  DEugenio2026,
Ji2025, 
Sun2026arXiv260120929S, 
Torralba2026A&A...707A..75T}.

As the system evolves, the (satellite) halo hosting the collapsed object may merge with or fall into the more massive galaxy \citep[e.g.,][step 3 in Figure~\ref{fig:cartoon}]{Visbal2014,Lupi_2021,Volonteri2025}. Some LRDs may be observed during an early phase in which the collapsing object remains offset from the star-forming galaxy, while others may correspond to later stages in which the halos have merged and the accreting source resides within the growing host galaxy, while still appearing red. 
As the dense gas is dispersed, the system may transition into a more typical galaxy–AGN configuration (step 4 in Figure~\ref{fig:cartoon}), potentially connecting to compact blue sources, the so-called “little blue dots” \citep[LBDs;][]{Brazzini2026arXiv260122214B, MadauMaiolino2026arXiv260222386M}. 

This evolutionary picture further provides an important connection between the early growth of the black hole and the buildup of the host galaxy. In this scenario, systems are initially born with a massive black hole and minimal stellar mass (if any), placing them much above the local $M_{\rm BH}$–$M_\star$ relation 
\citep[e.g.,][]{Furtak2024,Juodzbalis2024,Li2025ApJ...981...19L_overmassive}. As the black hole grows and the system subsequently merges with its companion (with typical stellar masses of $\sim10^8$–$10^9 M_\odot$), the system shifts downward in this plane.  
 

Depending on timing and subsequent star formation or accretion, the observable appearance of such systems should change over time (see also Section~\ref{sec:metallicity_sizes}) as the relative contributions of accretion-powered emission and stellar light vary, consistent with the diverse properties reported for the LRD population in recent studies \citep[e.g.,][]{Natarajan2017ApJ...838..117N,PerezGonzalez2024A,PerezGonzalez2026arXiv260220247P,Barro2025arXiv251215853B,Barro2026ApJ...997...48B,Leung2025,Billand2026A&A...706A..29B,Davis2026arXiv260223310D,HerreroCarrion2026,Sun2026arXiv260120929S}.
This evolutionary pathway also naturally impacts multiple observables, including the companion fraction (Section~\ref{sec:discussion:companionfraction}) and the metallicity and sizes of the red components (Section~\ref{sec:metallicity_sizes}), which pose apparent challenges to a simple direct-collapse picture but may be naturally explained by a combination of evolutionary effects and observational limitations, as we discuss below.

\begin{figure*}[t]
    \centering
     \includegraphics[width=0.9\linewidth]{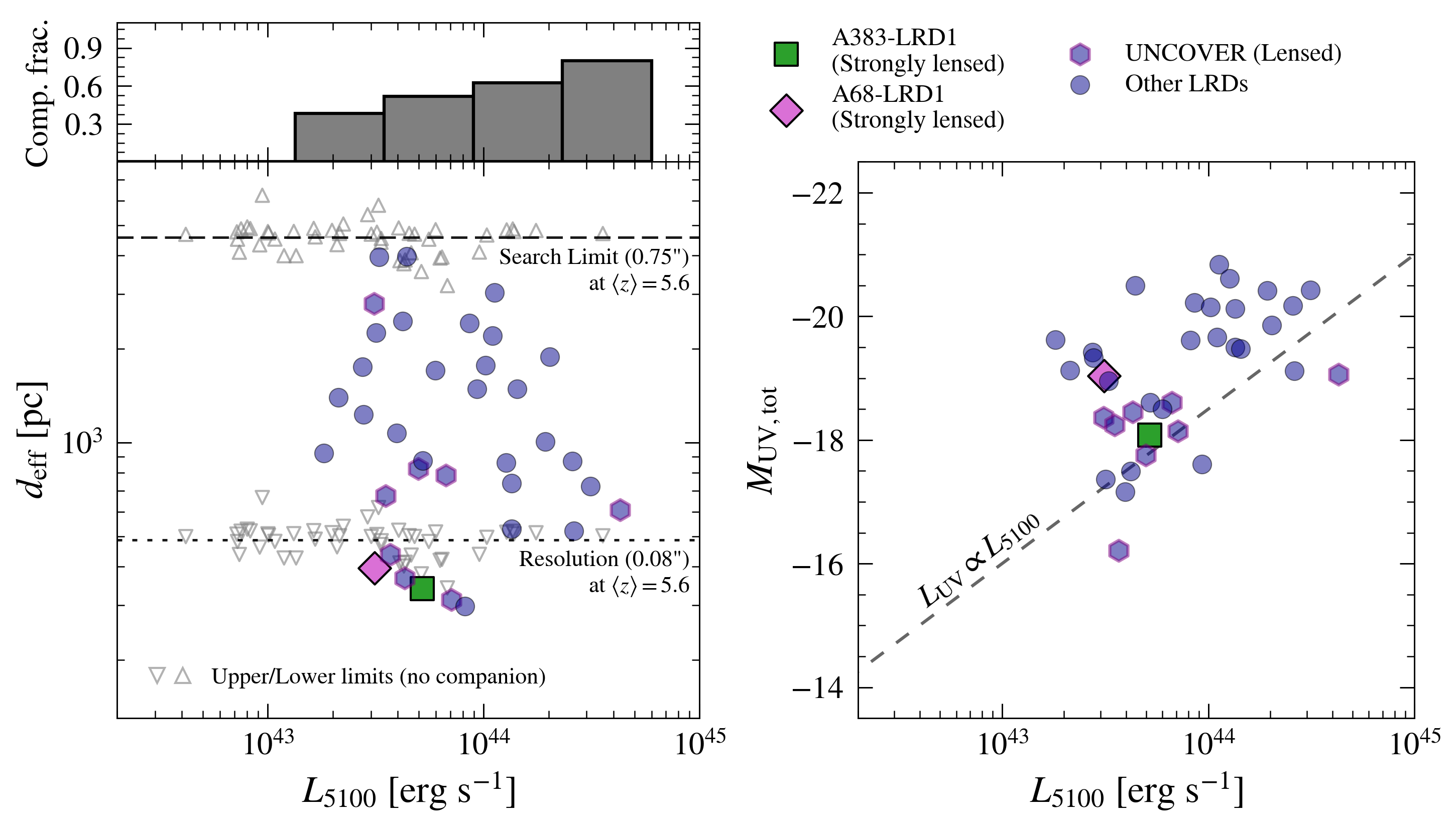} 
    \caption{\textit{Left panel}: Effective projected separation between the compact red component and its associated companion(s) as a function of the inferred luminosity $L_{5100}$.  The luminosity is measured using the NIRCam filter closest to rest-frame $5100\,\AA$ and converted to a luminosity using the source redshift. 
    The top panel shows the companion fraction as a function of luminosity; we find a striking trend where the companion fraction increases with $L_{5100}$, reaching nearly $\geqq$80-100\% completeness for the most luminous LRDs in our sample. 
    \textit{Right panel:} UV magnitude of the companion(s) (rest-frame 1500$\rm \AA$) as a function of  $L_{5100}$, indicating that companions associated with these high-luminosity LRDs are themselves significantly brighter in the UV.
   }
    \label{fig:sep_lum}
\end{figure*}

\subsection{Frequency of UV-Bright Companions}
\label{sec:discussion:companionfraction}
A strong prediction of this scenario is that every LRD has, or had at some point in its evolution, a companion object whose LW radiation exceeds the limit for H$_2$ dissociation. The observed fraction is high but not 100\,\% (see Section~\ref{sec:componentidentification}), seemingly in conflict with this expectation. 
However, the true fraction is almost certainly significantly higher than our reported fraction. 

First, our search is intentionally conservative, using $1\arcsec5$ cutouts and thus effectively focusing on companions $\sim$$0\arcsec75$ away from the compact red component. At the average redshift of the sample ($z=5.6$), this corresponds to a projected physical scale of $\sim$4.5~kpc. As illustrated in Fig.~\ref{fig:sep_lum}, for systems classified as lacking companions, the inferred projected distances should therefore be regarded as lower limits, as genuine physical companions may exist at larger separations. Establishing physical association at these distances generally requires wider-area searches and, ideally, spectroscopic confirmation. 

Second, some candidate companions are not included in the formal sample. We mark these features with question marks in Fig.~\ref{fig:RGB_selection_blue}. 
They fall into two distinct categories.
One class consists of companions at very small projected separations that are unresolved in unlensed fields. Strongly lensed systems demonstrate that companions can exist at separations that would be extremely difficult to resolve without lensing \citep{Baggen2025}. As companions are typically much fainter than the LRD, they are difficult to identify if they lie within $\sim2$ long-wavelength pixels. Adopting 2 pixels as an effective resolution limit corresponds to a physical scale of $\sim500$ pc, as indicated in Fig.~\ref{fig:sep_lum}. Systems classified as isolated may therefore host unresolved blue companions at smaller separations. Indeed, in several cases we observe faint blue residual features near the LRD position after subtraction of the red component (Fig.~\ref{fig:RGB_selection_blue}). 
These features likely correspond to intrinsically small three-dimensional separations ($r_{\rm 3D}\lesssim500$ pc). A small fraction of apparent overlaps could also arise from projection effects, where companions at larger separations happen to lie close to the line of sight. A simple geometric estimate shows that the probability of two sources appearing coincident in projection purely due to such alignment is small: for a characteristic red component size of $\sim100$ pc and a typical intrinsic separation of 1 kpc, the probability of such an overlap is only 1.5\%. 
Projection effects could therefore account for at most a small number of cases in the present sample.
The second class consists of sources that are clearly detected at larger projected separations and exhibit morphologies consistent with high-redshift star-forming galaxies, but for which the photometric redshift solutions do not satisfy our statistical consistency criterion. This typically reflects limited band coverage or low SNR rather than contradictory redshift constraints. Although these sources are visually compelling and consistent with being physical companions, we exclude them from the formal sample. 
Including all such candidate companions marked with question marks in Fig.~\ref{fig:RGB_selection_blue} would increase the inferred companion fraction to 72\%.

Third, the data are of mixed depth, and very faint companions are likely missed in our analysis,
particularly at higher redshifts ($z\gtrsim 7$).
As the LW flux scales with $d_{\rm eff}^{-2}$, even very faint companions can produce sufficient radiation to bring the LRD into the direct-collapse regime if they are sufficiently close. 
In our sample the UV luminosity of the companions and the fraction of LRDs with a detected companion both scale with the $L_{5100}$
luminosity of the LRD (Fig.~\ref{fig:sep_lum}), strongly suggesting that depth effects plague the
faint end of the sample. 
If we limit the sample to higher luminosities, the companion fraction increases, reaching $\sim$80\% in the highest-luminosity bin (Figure~\ref{fig:sep_lum}).
Consistent with this interpretation, LRDs in lensed fields, such as the UNCOVER sample, show a substantially higher companion fraction than those in unlensed fields, reflecting the improved sensitivity to faint, close companions.


Fourth, the companion may no longer be observable as a distinct object. The star-forming phase required to generate a high LW flux may be short-lived, such that some systems are observed after the companion has faded or been disrupted \citep{Schaerer_2002}, or after it has merged with the host galaxy (step 3 in Figure \ref{fig:cartoon}). 
An interesting case in this context is UNCOVER-20466, one of the LRDs in our sample without a clearly detected UV-bright companion. Recent JWST/NIRSpec IFU observations reveal evidence for a dense stellar component hosting an overmassive black hole \citep{Jones2026MNRAS.546ag115J}. This LRD may represent a later evolutionary stage in which the originally distinct companion has merged. However, given its high redshift ($z=8.5$), it is also possible that any close companion remains difficult to separate or falls below the detection limits, as discussed earlier in this section.


Finally, the LRD selection does not require that the observed emission in all systems be powered by (exotic) accreting black holes.
A fraction of the population may instead be dominated by compact stellar components, given that the emission from stellar populations in at least a subset of the sources remains largely degenerate with accretion-dominated emission, especially at low metallicities and high ionization parameters  \citep{Cleri2025ApJ...994..146C}. Such stellar-dominated systems can reach high central stellar densities and give rise to broad emission lines without requiring an actively accreting black hole \citep{Baggen2024, Guia2024}.
We should expect ``normal" stellar Balmer breaks, consistent with the presence of evolved stellar populations \citep{Wang2024_balmer,Weibel2025ApJ...983...11W_quiescentz7}. 
Such sources likely represent the earliest phases of 
dense stellar cores that will evolve into the central regions of massive galaxies at later times  \citep[e.g.,][]{Labbe2023,Baggen2023,Baggen2024,Guia2024,Rinaldi2025arXiv250717738R,HerreroCarrion2026}. The extremely high stellar densities reached in these systems may nevertheless rapidly lead to the formation of massive black hole seeds through runaway stellar collisions or dynamical processes \citep[e.g.,][]{Omukai2008,Tagawa_2020,Escala2025ApJ...995...44E,Pacucci2025,Vergara2026A&A...707A..71V,Vergara2025A&A...704A.321V}.
The fraction of LRDs that are stellar-dominated remains uncertain. Although a subset resides in overdense environments \citep{Tanaka2024arXiv241214246T_dualLRD, Matthee2025ApJ...988..246M,Schindler2025NatAs...9.1732S, DEugenio2026,Torralba2026}, the clustering and number density of LRDs indicate that most reside in relatively low mass halos \citep{Carranza-Escudero2025,Pizzati2025MNRAS.539.2910P}.

\subsection{Metallicities and Sizes}
\label{sec:metallicity_sizes}
In classical theoretical models of direct collapse it is assumed that the gas has near-zero metallicity such that H$_2$ cooling is the dominant channel. 
Above a critical metallicity of order $\sim10^{-5}\,Z_\odot$, metal-line cooling is expected to induce fragmentation \citep{Begelman1978MNRAS.185..847B}.
 The scenario described here thus
predicts that LRDs had near-zero metallicities during collapse.

Observationally, this expectation appears to be in tension with the properties of at least part of the LRD population.
Many LRDs show prominent metal emission lines \citep[e.g.,][]{Labbe2024, DEugenio2025, Kokorev2025arXiv251107515K,Lambrides2025arXiv250909607L,Torralba2026A&A...707A..75T}, indicating that at least some chemical enrichment has occurred in these systems
(although the enriched gas could also come from the UV-bright companion, or other gas in the vicinity).
At the same time, other LRDs show evidence for low metallicity \citep{Maiolino2025_pristine, Tripodi2025NatCo..16.9830T},
with some LRD spectra dominated by hydrogen and helium lines and little to no detectable metal emission \citep{Graaff2025_Cliff, Furtak2024}.

A related prediction concerns sizes. The immediate products of collapse, such as supermassive stars (in self-gravitating disks), quasi-stars, or “black hole stars,” are expected to be very small,  with characteristic scales ranging from subparsec to a few parcsecs \citep[e.g.,][]{Pacucci2026arXiv260114368P},
and thus unresolved by JWST, even when strongly lensed. 
In contrast, we measure a wide range of effective sizes for the red components, with a mean effective radius of $\sim$200~pc, consistent with stacked rest-frame optical sizes reported by \citet{Zhang2025}.

Combined, both the observed metallicities and the range of measured sizes can be naturally reconciled if some star formation and associated enrichment occurred during or after the initial collapse, or during subsequent interactions with nearby companions 
\citep[step 3 in Figure~\ref{fig:cartoon}; see also][]{Natarajan2017ApJ...838..117N}. 
Such evolution would lead to sizes larger than expected for a single compact object, and, given the extreme compactness of these systems, enrichment is expected to proceed rapidly. In this picture, signatures of massive stars may be present in the spectra of the red components \citep[e.g., nitrogen emission;][]{Berg2025, Morel2025arXiv251120484M}, particularly in spatially resolved or more metal-rich systems.

It may also be that the collapse does not proceed in a strictly classical, pristine manner. In metal-poor gas exposed to strong LW radiation fields, growth may remain dominated by so-called ``supercompetitive accretion", in which gas is preferentially funneled toward the most massive object, allowing one or more massive seeds to form despite the presence of additional fragments \citep[e.g.,][]{Chon2020MNRAS.494.2851C}. This process can operate up to metallicities of $Z \lesssim 10^{-3}\,Z_\odot$, indicating that the requirement of strictly pristine gas may be relaxed~\citep{Chon2020MNRAS.494.2851C,Regan_2020}. This may naturally lead to configurations with multiple compact components \citep[e.g.,][]{Latif2020ApJ...892L...4L_binary,Woods2024ApJ...960...59W}. 
Interestingly, \citet{Yanagisawa2026} report a system (``Red Eyes'') containing two compact red components separated by $\sim$70~pc, which may be consistent with such a scenario.


\section{Conclusions}
In this work, we have identified candidate UV-bright companions to a substantial fraction of LRDs (43\% increasing to 72\% when including tentative companions), assumed a physical connection between the components, and used their luminosities and projected separations to infer the local LW radiation fields experienced by the compact red sources. The inferred radiation fields are sufficiently intense to plausibly suppress molecular cooling and enable rapid collapse to very high densities, in line with established theoretical models. While this agreement does not constitute a proof of the scenario, it demonstrates that the picture is internally consistent and observationally viable.

It is particularly striking that the two most strongly lensed systems in our sample exhibit remarkably similar configurations, each consisting of an extremely compact blue component located only $\sim$300~pc from a compact red source. This recurring geometry suggests that the association may not be coincidental. In the framework explored here, the UV-bright component exists first and regulates the formation of the compact red object, though alternative scenarios in which the red component forms first and subsequently triggers nearby star formation remain plausible. The present data do not uniquely determine the direction of causality, but they motivate a unified picture that naturally connects the observed morphologies, SEDs, and inferred radiation fields.

This concordance picture leads to a set of concrete, testable predictions and motivates theoretical and observational follow-up studies.
An important question is whether the observed abundance of LRDs is consistent with theoretical expectations for objects formed through LW-regulated collapse. 
Early work used cosmological N-body simulations to obtain the number density of close synchronized pairs and found up to a few $\times10^{-4}$ cMpc$^{-3}$ at redshift $z{\sim}10$ \citep{Visbal2014}. Recent observational estimates suggest LRD number densities of $\sim10^{-4.5}$ cMpc$^{-3}$  $z>4$ \citep{Ma2026ApJ..1000...59M}. Adopting this value and taking 40\%, as a conservative lower limit on the fraction of LRDs with close UV-bright companions implies a number density of $\sim10^{-4.9}$ cMpc$^{-3}$. 
At face value, this is somewhat lower than early theoretical predictions. However, these theoretical predictions apply to higher redshift, and the expected abundance may decline toward lower redshift as suitable pristine environments become rarer. More recent work based on cosmological hydrodynamical simulations and analytic prescriptions also finds somewhat lower maximum number densities \citep{OBrennan+2025}. In addition, recent simulations indicate that both the number density and redshift evolution of DCBH models are consistent with the observed LRD population \citep{Jeon2026ApJ...998..148J}.
 Whether current models quantitatively reproduce the observed number density properties of LRDs therefore remains an open question.  
A further prediction of the direct-collapse scenario is that such systems should reside in highly biased environments. Measurements of the clustering of LRDs therefore provide an important independent test of this formation pathway \citep{Wang2026arXiv260315736W_clustering}.
 In addition, simulations of the collapse process and subsequent star formation and enrichment will be highly valuable.

On the observational side, we can expect progress
from spatially resolved spectroscopy, ideally for strongly lensed systems such as A383-LRD1 and A68-LRD1.
Such observations could measure metallicity and ionization gradients, constrain the degree of radiative shielding between components, and provide accurate sizes (or size limits) of the red components. These data will determine whether the close proximity of UV-bright companions to compact red sources reflects a causal link in their formation or requires an alternative physical origin.

\begin{acknowledgments}
We thank Earl Bellinger, Fabio Pacucci, Andrea Ferrara and Dale Kocevski for useful discussions. 
This work is based on observations made with the NASA/ESA/CSA James Webb Space Telescope. The data were obtained from the Mikulski Archive for Space Telescopes at the Space Telescope Science Institute, which is operated by the Association of Universities for Research in Astronomy, Inc., under NASA contract NAS 5-03127 for JWST. These imaging observations are associated with programs 
1345, 1180, 1181, 1243, 6882, 2561, 1324, 4111, 1895. 
The compiled dataset can be accessed at \url{https://doi.org/10.17909/1m8f-9c47}. 
The Cosmic Dawn Center (DAWN) is funded by
the Danish National Research Foundation under grant DNRF140. J.M. and A.T. acknowledge funding by the European Union (ERC, AGENTS,  101076224).
This work was performed in part at Aspen Center for Physics, which is supported by National Science Foundation grant PHY-2210452.
This work used the following Python packages: \texttt{Matplotlib} \citep{Matplotlib_Hunter2007}, \texttt{SciPy} \citep{SciPy_Virtanen2020}, \texttt{NumPy} \citep{numpy_vanderWalt2011}, \texttt{AstroPy} \citep{Astropy2022} and \texttt{colossus} \citep{Diemer2018_colossus} and \texttt{photutils} \citep{larry_bradley_2025_14889440}.

\end{acknowledgments}


%

\clearpage
\appendix
\section{Sources with no identified companion}
\label{app:sources_no_comp}
Table~\ref{tab:appendix_table}
shows the list of LRDs without a detected companion.
\begin{table}[h!]
    \centering

    \begin{minipage}[t]{0.42\textwidth}
        \centering
        \small
        \begin{tabular}{lccc}
        \hline
        ID & RA & DEC & $z$ \\
        \hline
        CEERS-3153 & 214.9258 & 52.9457 & 5.09 \\
        CEERS-5760 & 214.9724 & 52.9622 & 5.08 \\
        CEERS-6126 & 214.9234 & 52.9256 & 5.29 \\
        CEERS-7000 & 214.8400 & 52.8606 & 3.86 \\
        CEERS-12833 & 214.9295 & 52.8879 & 7.16 \\
        CEERS-13135 & 214.8868 & 52.8554 & 4.95 \\
        CEERS-13748 & 214.8405 & 52.8179 & 6.11 \\
        CEERS-14949 & 214.9910 & 52.9165 & 5.68 \\
        CEERS-18850 & 214.9278 & 52.8500 & 7.48 \\
        CEERS-20496 & 215.0783 & 52.9485 & 6.79 \\
        CEERS-20777 & 215.1371 & 52.9886 & 5.29 \\
        CEERS-23931 & 214.8171 & 52.7483 & 9.94 \\
        CEERS-69459 & 214.8897 & 52.8330 & 5.67 \\
        CEERS-80072 & 214.9372 & 52.9654 & 5.30 \\
        CEERS-81443 & 214.8307 & 52.8878 & 7.76 \\
        CEERS-82815 & 214.8091 & 52.8685 & 5.62 \\
        CEERS-99879 & 214.8719 & 52.8804 & 5.68 \\
        CEERS-111399 & 214.8968 & 52.8758 & 6.13 \\
        CEERS-114801 & 214.9755 & 52.9253 & 5.11 \\
        CEERS-19300 & 215.0221 & 52.9208 & 4.53 \\
        CEERS-23961 & 214.8997 & 52.8128 & 5.00 \\
        CEERS-2015 & 214.9180 & 52.9372 & 4.89 \\
        CEERS-47962 & 214.8925 & 52.8569 & 6.73 \\
        UNCOVER-2008 & 3.5924 & -30.4328 & 6.74 \\
        UNCOVER-13123$^\dagger$ & 3.5798 & -30.4016 & 7.04 \\
        UNCOVER-15383$^\dagger$ & 3.5835 & -30.3967 & 7.04 \\
        UNCOVER-16594$^\dagger$ & 3.5972 & -30.3943 & 7.04 \\
        UNCOVER-20466 & 3.6404 & -30.3864 & 8.50 \\
        UNCOVER-28876 & 3.5696 & -30.3732 & 7.04 \\
        UNCOVER-20698 & 3.5567 & -30.4082 & 2.42 \\
        UNCOVER-27114 & 3.5458 & -30.3957 & 5.20 \\
        UNCOVER-49702 & 3.5477 & -30.3337 & 4.87 \\
        \hline
            \end{tabular}
        \end{minipage} 
        \hspace{2mm}%
    \begin{minipage}[t]{0.42\textwidth}
        \centering
        \small
        \begin{tabular}{lccc}
        \hline
        ID & RA & DEC & $z$ \\
        \hline
        JADES-39353 & 189.2940 & 62.1531 & 4.85 \\
        JADES-4685 & 189.0963 & 62.2391 & 7.42 \\
        JADES-12402 & 53.1327 & -27.7655 & 3.19 \\
        JADES-38562 & 53.1359 & -27.8716 & 4.82 \\
        JADES-4461 & 53.2040 & -27.7721 & 7.25 \\
        JADES-12068 & 53.1265 & -27.8181 & 5.92 \\
        JADES-21598 & 53.0877 & -27.8712 & 4.74 \\
        JADES-64894 & 53.0605 & -27.8484 & 5.50 \\
        GOODS-N-4014 & 189.3001 & 62.2120 & 5.23 \\
        GOODS-N-12839 & 189.3443 & 62.2634 & 5.24 \\
        GOODS-N-13733 & 189.0571 & 62.2689 & 5.24 \\
        GOODS-N-14409 & 189.0721 & 62.2734 & 5.14 \\
        GOODS-N-15498 & 189.2855 & 62.2808 & 5.09 \\
        GOODS-N-16813 & 189.1793 & 62.2925 & 5.36 \\
        J1148-18404 & 177.0580 & 52.8628 & 5.01 \\
        J1120-7546 & 169.9994 & 6.6547 & 4.97 \\
        J1120-14389 & 170.0037 & 6.7196 & 4.90 \\
        \hline
        \end{tabular}
    \end{minipage}
     \caption{Sources with no identified companion. $\dagger$ A2744 triply lensed LRD: Although a potential companion is identified by the pipeline, we remain cautious. For a genuinely associated companion, all three lensed images are expected to show consistent counterparts. Proper delensing to the source plane using accurate lens models, together with spectroscopic confirmation, is required to assess this. We therefore do not count this object as a reliable companion in our analysis. }
     \label{tab:appendix_table}
\end{table}

\section{A strongly lensed Little Red Dot behind Abell 68}
\label{app:a68_triple}
\begin{figure}
    \centering
    \includegraphics[width=0.8\linewidth,  trim={0cm 0cm 0cm 0.0cm}, clip]{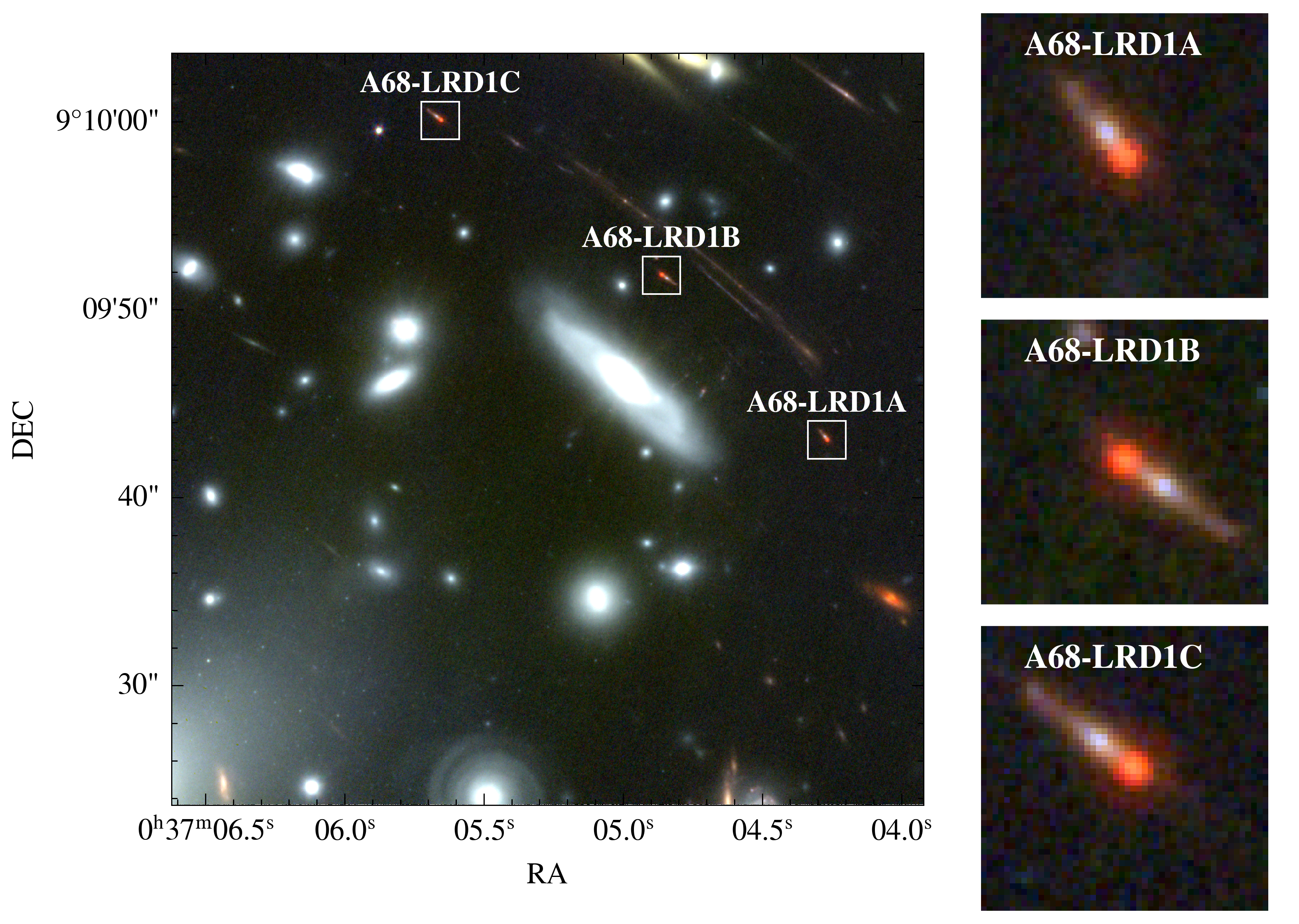}
    \caption{RGB composite 
    of the A68 cluster field, constructed from F444W+F356W (red), F277W+F200W (green), and F150W+F115W+F090W (blue),
    showing the three lensed images of A68-LRD1. The imaging is publicly available from the VENUS collaboration (Program ID: GO~6882; PI: S.~Fujimoto).
    }
    \label{fig:abell_68_cutouts}
\end{figure}

As part of our search for strongly lensed systems, we identified a new strongly LRD, triply imaged behind the galaxy cluster A68, observed as part of the VENUS collaboration (Program ID: GO~6882; PI: S.~Fujimoto). It is visually identified and corresponds to the triply imaged source reported by \citet{Richard2007}, who measured a spectroscopic redshift of $z=5.421$ from the Ly$\alpha$ emission line.
We do not construct new lens models for this system, but instead adopt the magnification estimates from \citet{Richard2007}, namely $\mu_1 = 12.5 \pm 0.9$, $\mu_2 = 14.3 \pm 0.9$, and $\mu_3 = 11.9 \pm 1.1$.
Fig \ref{fig:abell_68_cutouts} shows the RGB composite image of the galaxy cluster and the three lensed images of A68-LRD1.
The morphology is remarkably similar to the LRD found behind A383 \citep{Baggen2025, Golubchik2025}, motivating the concordance picture developed in this work. In this analysis, we perform the measurements on image A68-LRD1A.

\clearpage
\bibliography{sample701}{}
\bibliographystyle{aasjournalv7}



\end{document}